\documentclass[11pt,prd,preprintnumbers,amsmath,amssymb,nofootinbib]{revtex4}

\pdfoutput=1
\usepackage{paralist}
\usepackage{graphicx}
\usepackage{enumitem}
\usepackage{bm}
\usepackage{color}

\newcommand{\eq}[1]{Eq.~(\ref{#1})}

\newcommand{\bea} {\begin{eqnarray}}
\newcommand{\eea} {\end{eqnarray}}
\newcommand{\ba} {\begin{eqnarray*}}
\newcommand{\ea} {\end{eqnarray*}}

\begin{document}

\preprint{ICCUB-16-031, PSI-PR-16-12}

\title{Loop effects of heavy new scalars and fermions in $b\to s\mu^+\mu^-$ }
\author{Pere Arnan}
\email{arnan@ub.edu}
\affiliation{Departament de F\'{\i}sica Qu\`antica i Astrof\'{\i}sica (FQA),
Institut de Ci\`encies del Cosmos (ICCUB), Universitat de Barcelona (UB), Spain}
\author{Andreas Crivellin}
\email{andreas.crivellin@cern.ch}
\affiliation{Paul Scherrer Institut, CH--5232 Villigen PSI, Switzerland}
\author{Lars Hofer}
\email{hofer@fqa.ub.edu}
\affiliation{Departament de F\'{\i}sica Qu\`antica i Astrof\'{\i}sica (FQA),
Institut de Ci\`encies del Cosmos (ICCUB), Universitat de Barcelona (UB), Spain}
\author{Federico Mescia}
\email{mescia@ub.edu}
\affiliation{Departament de F\'{\i}sica Qu\`antica i Astrof\'{\i}sica (FQA),
Institut de Ci\`encies del Cosmos (ICCUB), Universitat de Barcelona (UB), Spain}

\begin{abstract} 
\vspace{20mm}
Recent measurements of $b\to s\mu^+\mu^-$ processes at LHCb and BELLE have revealed tensions at the $2-3\,\sigma$ level
between the Standard Model (SM) prediction and the experimental results in the channels $B\to K^*\mu^+\mu^-$ and $B_s\to\phi\mu^+\mu^-$, as well as in the lepton-flavor universality violating observable $R_K={\rm Br}(B\to K\mu^+\mu^-)/{\rm Br}(B\to Ke^+e^-$). Combined global fits to the available $b\to s\mu^+\mu^-$ data suggest that these tensions might have their common origin in New Physics (NP) beyond the SM because some NP scenarios turn out to be preferred over the SM by $4-5\,\sigma$. The fact that all these anomalies are related to muons
further suggests a connection (and a common NP explanation) with the long-standing anomaly in the anomalous magnetic moment of the muon, $a_\mu$. In this article, we study the impact of a generic class of NP models featuring new heavy scalars and fermions that couple to the SM fermions via Yukawa-like interactions. We consider two different scenarios, introducing either one additional fermion and two scalars or two additional fermions and one scalar, and examine all possible representations of the new particles under the SM gauge group with dimension up to the adjoint one. The models induce one-loop contributions to $b\to s\mu^+\mu^-$ and $a_\mu$ which are capable of solving the respective anomalies at the $2\sigma$ level, albeit a relatively large coupling of the new particles to muons is required.  In the case of $b\to s\mu^+\mu^-$, stringent constraints from $B_s-\overline{B}_s$ mixing arise which can be relaxed if the new fermion is a Majorana particle.
\end{abstract} \maketitle

\newpage 

\section{Introduction}
While a direct production of particles beyond the {ones of the SM} has not been observed at the LHC so far, some observables in the flavor sector show tensions {with the theory predictions} that can be interpreted as indirect 'hints' for new physics. {The} affected channels/observables comprise $B\to K^* \mu^+\mu^-$,
$B_s\to\phi\mu^+\mu^-$ and $R_K={\rm Br}[B\to K \mu^+\mu^-]/{\rm Br}[B\to K e^+e^-]$, 
all of them induced by the same quark-level transition $b\to s\mu^+\mu^-$~\footnote{Deviations from SM predictions were
also observed in tauonic $B$ decays. Since these tensions cannot be 
explained by loop effects, we do not discuss them in this article.}.

Let us give a brief account on the experimental and theoretical situation concerning
$b\to s\mu^+\mu^-$ transitions. In the decay $B\to K^*\mu^+\mu^-$, tensions between 
the SM prediction and the LHCb data~\cite{LHCb:2015dla} mainly manifest themselves
as a $\sim 3\sigma$ anomaly in the angular observable $P_5^\prime$\cite{Descotes-Genon:2013vna,Matias:2012xw}. This observable is fairly robust with respect to hadronic uncertainties~\cite{Descotes-Genon:2014uoa} because,
at leading order in $\alpha_s$ and $\Lambda/m_B$, form factors cancel as a consequence of large-recoil symmetries~\cite{Kruger:2005ep}. Very recently, a (less precise) BELLE measurement~\cite{Abdesselam:2016llu} confirmed
the $P_5^\prime$ anomaly at the $2\sigma$ level. In the channel $B_s\to\phi\mu^+\mu^-$, the branching ratio 
measured by LHCb~\cite{Aaij:2015esa} in the region of large $\phi$-recoil is at $\gtrsim 2\sigma$ in conflict with the SM prediction based on light-cone sum-rule form factors~\cite{Straub:2015ica}. Finally, LHCb has observed  lepton flavor universality violation (LFUV) in $B\to K\ell^+\ell^-$ decays~\cite{Aaij:2014ora}: in the range $1\,{\rm GeV}^2<m_{\ell\ell}^2<6\,{\rm GeV}^2$ of the dilepton invariant mass $m_{\ell\ell}$, the measured ratio
$R_K$ deviates from the theoretically clean SM prediction~\cite{Bordone:2016gaq,Hiller:2003js} by $2.6\sigma$.  Global fits, including the above observables 
as well as other $b\to s$ data like $B_s\to\mu^+\mu^-$, $b\to s\gamma$, etc., found that scenarios with a new physics (NP) contribution to the operators \begin{equation}
 {O_9^{(\prime)}}=\frac{\alpha_\mathrm{EM}}{{4\pi}} \left[\bar s{\gamma^\nu }{P_{L(R)}}b\right] \left[\bar \mu {\gamma_\nu }\mu\right]\,,
 \hspace{2cm}{O_{10}^{(\prime)}}=\frac{\alpha_\mathrm{EM}}{{4\pi}}  \left[\bar s{\gamma^\nu
 }{P_{L(R)}}b\right] \left[\bar \mu {\gamma_\nu }{\gamma^5}\mu \right]
\end{equation}
can significantly improve the description of the data compared to the SM~\cite{Altmannshofer:2015sma,Descotes-Genon:2015uva,Hurth:2016fbr}. Depending on the underlying model of NP, a different pattern of correlations among the NP Wilson coefficients $C_9^{(\prime)},C_{10}^{(\prime)}$ arises, and for instance models that only generate $C_9$, $C_9=-C_{10}$ or $C_9=-C_{9}'$ with a large negative $C_9$ are preferred over the SM by $4$--$5\sigma$. 

At the level of concrete NP models, most analyses focus on a generation of the required NP effects at tree level, either by {the} exchange of $Z'$ vector bosons~\cite{Descotes-Genon:2013wba,Gauld:2013qba,Gauld:2013qja,Buras:2013dea,Altmannshofer:2014cfa,Crivellin:2015mga,Crivellin:2015lwa,Niehoff:2015bfa,Sierra:2015fma,Crivellin:2015era,Celis:2015ara,Allanach:2015gkd,Celis:2016ayl,Boucenna:2016wpr,Boucenna:2016qad,Megias:2016bde} or {through}
leptoquarks~\cite{Gripaios:2014tna,Belanger:2015nma,Becirevic:2015asa,Varzielas:2015iva,Alonso:2015sja,Calibbi:2015kma,Barbieri:2015yvd,Bauer:2015knc}.
An explanation of the anomalies {via} loop effects, on the other hand, typically leads to correlated imprints on other observables like {the} anomalous magnetic moment of the muon {($a_\mu$)}. It is thus appealing to investigate the possibility of a simultaneous solution of the flavor anomalies and the long-standing tension in $(g-2)_\mu$ at the loop level,  for example by light $Z^\prime$ bosons~\cite{Langacker:2008yv,Baek:2001kca,Ma:2001md,Gninenko:2001hx,Pospelov:2008zw,Heeck:2011wj,Biggio:2016wyy,Harigaya:2013twa,Altmannshofer:2014pba,Altmannshofer:2016brv}, leptoquarks~\cite{Chakraverty:2001yg,Cheung:2001ip,Bauer:2015knc} or new fermions and scalars~\cite{Iltan:2001nk,Omura:2015nja,Altmannshofer:2016oaq,Broggio:2014mna,Wang:2014sda,Abe:2015oca,Crivellin:2015hha,Batell:2016ove,Freitas:2014pua,
Gripaios:2015gra,Goertz:2015nkp} (see Ref.~\cite{Stockinger:2006zn} for a review on the situation in SUSY).

In this article, we examine in detail the possibility proposed in Ref.~\cite{Gripaios:2015gra} 
that the anomalies in the $b\to s\mu^+\mu^-$ data and $(g-2)_\mu$ are explained by loop effects
involving heavy new scalars and fermions that couple to the SM fermions via Yukawa-like 
interactions. In order to generate the Wilson coefficient $C_9$, the new particles must
couple to the left-handed SM quark doublets $Q$. We study the minimal setup in which 
the new particles do not couple to right-handed SM fields, implying $C_9=-C_{10}$ which 
is one of the favored patterns for the solution of the $b\to s\mu^+\mu^-$ anomalies.
Whereas the emphasis in Ref.~\cite{Gripaios:2015gra} was on model-building aspects 
and a particular higher-dimensional representation for the new particles under the SM gauge group, 
we explore in more detail the phenomenological consequences in {a} general class of models{:}
We consider those representations which are realized in the SM (singlet, fundamental and adjoint) and 
study also the case of the fermions {(scalars)} being Majorana particles {(real scalars)}.

The paper is organized as follows: In Sec.~\ref{sec:setup} we define {our} model and classify the various
representations under the SM gauge group for the new particles. In Sec.~\ref{sec:observables} we give the formulae for the Wilson coefficients and the observables relevant for our numerical analysis in Sec.~\ref{sec:numerics}. Finally we conclude in Sec.~\ref{sec:conclusions}.

\section{Setup}
\label{sec:setup}

\begin{figure*}[t]
\begin{center}
\begin{tabular}{cp{7mm}c}
\includegraphics[width=0.9\textwidth]{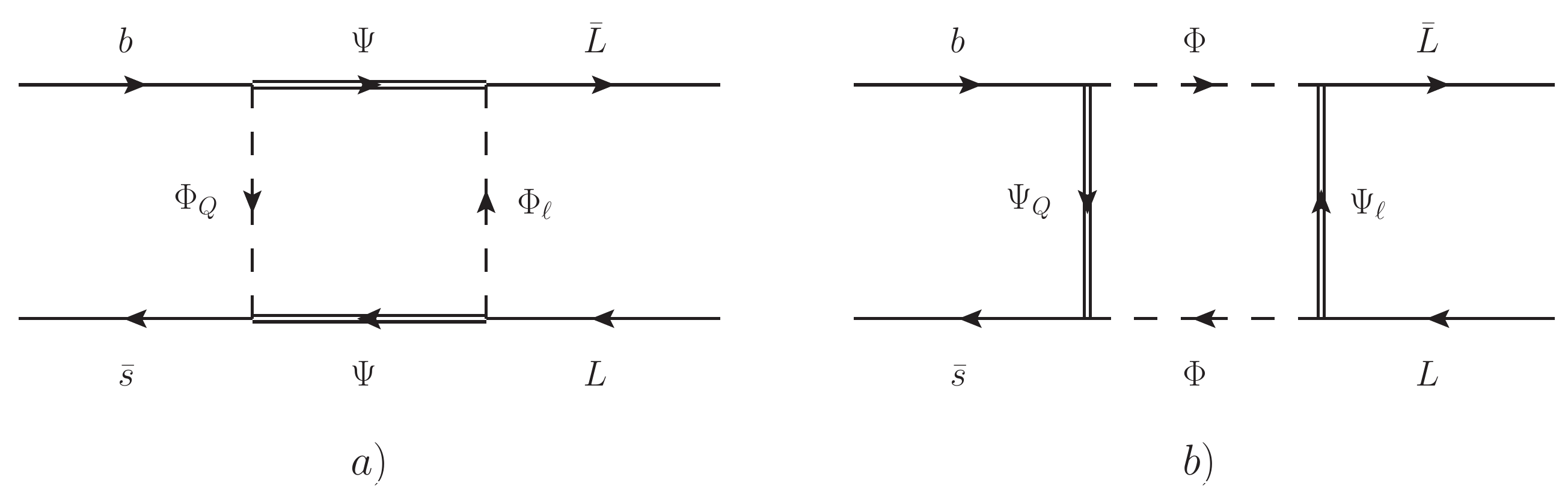}
\end{tabular}
\end{center}
\caption{Box-diagrams contributing to $b\to s \ell^+\ell^-$ and  $b\to s \nu\bar\nu$ in case $a)$ and $b)$.
\label{fig:Boxes} }
\end{figure*}

In the spirit of Ref.~\cite{Gripaios:2015gra}, we introduce new heavy scalars and vector-like fermions in such a way that a 
one-loop box contribution to $b\to s\mu^+\mu^-$ is generated (see Fig.~\ref{fig:Boxes}). As mentioned in the introduction,
we will assume that the new particles only couple to left-handed SM fermions. This assumption minimizes the number of free 
parameters and is phenomenologically well motivated because the resulting pattern $C_9=-C_{10}$ is one of the scenarios that are suited best for the description of $b\to s\mu^+\mu^-$ data. To draw the diagram on the left-hand side of Fig.~\ref{fig:Boxes}, we need a new fermion $\Psi$ that couples to both quarks and leptons, and two different scalar
particles (with different color quantum numbers), one of them coupling to quarks and one of them to leptons. Alternatively,
exchanging the roles played by the fermions and scalars, we get the diagram on the right-hand side of Fig.~\ref{fig:Boxes}.
Therefore, we construct the following two distinct models:
\begin{enumerate}[label=\alph*)]
\item One additional fermion $\Psi$ and two additional scalars $\Phi_Q$ and $\Phi_\ell$ with interactions described by the Lagrangian
\begin{eqnarray}
{\cal L}_{{\mathop{\rm int}} }^{a)} =  {\Gamma^Q_i\bar Q_i }{P_R}{\Psi}{\Phi_Q} + \Gamma^L_i \bar L_i {P_R}{\Psi}{\Phi_\ell} +
{\rm{h}}{\rm{.c.}}.
\label{eq:Linta}
\end{eqnarray}
\item Two additional fermions $\Psi_Q$ and $\Psi_\ell$ and one additional scalar $\Phi$ with interactions described
by the Lagrangian
\begin{eqnarray}
{\cal L}_{{\mathop{\rm int}} }^{b)} =  {\Gamma^Q_i\bar Q_i}{P_R}{\Psi_Q}{\Phi} + \Gamma^L_i \bar L_i {P_R}{\Psi_\ell}{\Phi} +
{\rm{h}}{\rm{.c.}}\,.
\label{eq:Lintb}
\end{eqnarray}
\end{enumerate}
In Eqs.~(\ref{eq:Linta}) and (\ref{eq:Lintb}), $Q_i$ and $L_i$ denote the left-handed quark and lepton 
doublets with family index $i$. The box-diagrams contributing to $b\to s \ell^+\ell^-$ and $b\to s \bar \nu\nu$ for the model classes a) and b) are {shown} in Fig.~\ref{fig:Boxes}. Analogous box diagrams induce $B_s-\overline{B}_s$ mixing (see upper row in Fig.~\ref{fig:diagrams_Bs-mixing}).

\begin{figure*}[t]
\begin{center}
\begin{tabular}{cp{7mm}c}
\includegraphics[width=0.9\textwidth]{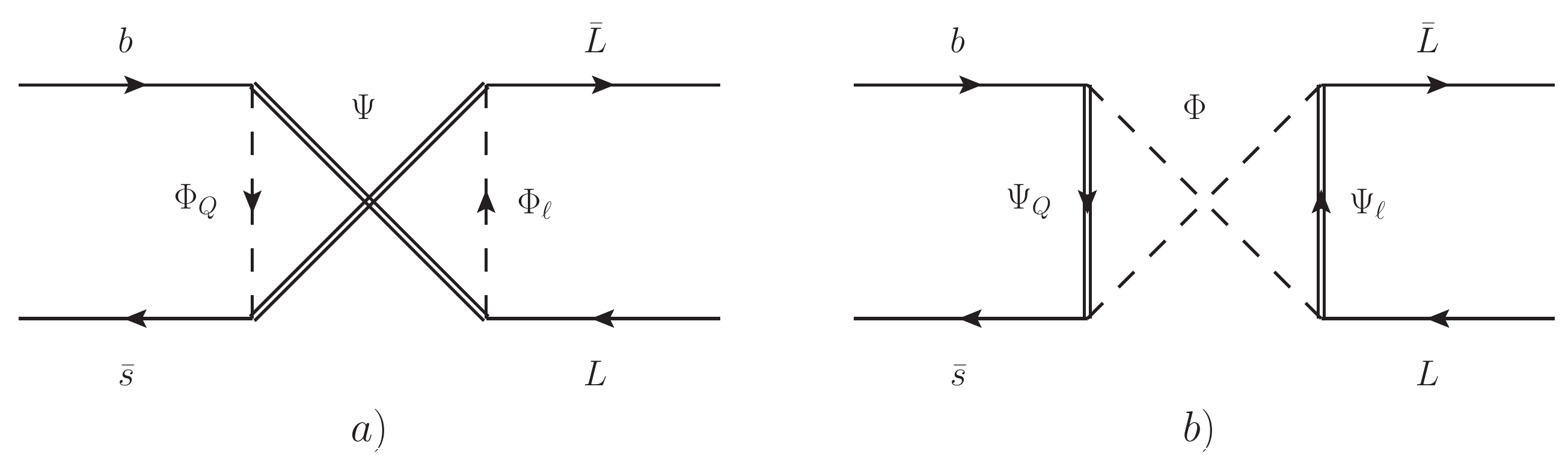}
\end{tabular}
\end{center}
\caption{Crossed boxes contributing to $b\to s \ell^+\ell^-$ and  $b\to s\nu\bar\nu$ in case $a)$ ($b$)) if $\Phi$ ($\Psi$) is a Majorana fermion (real scalar).
\label{fig:Boxescrossed} }
\end{figure*}

One-loop contributions to $b\to s\ell^+\ell^-$ can also be generated by the crossed box diagrams shown in Fig.~\ref{fig:Boxescrossed}. Whereas the standard box contributions in Fig.~\ref{fig:Boxes} derive from the Lagrangian ${\cal L}_{{\mathop{\rm int}} }^{a)}$ $\left({\cal L}_{{\mathop{\rm int}} }^{b)}\right)$ with $\Psi$ ($\Phi$) coupling both to quarks and to leptons, crossed boxes are induced by a variant ${\cal L}_{{\mathop{\rm int}} }^{a^\prime)}$ (${\cal L}_{{\mathop{\rm int}} }^{b^\prime)}$) of the Lagrangian where $\Psi$ ($\Phi$) couples to quarks and the charge-conjugated field $\Psi^c$ ($\Phi^c$) to leptons. Therefore, in the case of a Majorana fermion $\Psi=\Psi^c$ (neutral scalar $\Phi=\Phi^c$) one has ${\cal L}_{{\mathop{\rm int}} }^{a)}={\cal L}_{{\mathop{\rm int}} }^{a^\prime)}$ (${\cal L}_{{\mathop{\rm int}} }^{b)}={\cal L}_{{\mathop{\rm int}} }^{b^\prime)}$),
and both crossed and uncrossed boxes are present. In the other cases, it turns out that the primed Lagrangians lead to a very similar phenomenology\footnote{Predictions for observables involving only quarks or only leptons are identical for the primed and unprimed Lagrangians. For $b\to s\ell^+\ell^-$ the impact on our 
phenomenological analysis consists in a sign change of the Wilson coefficient $C_9=-C_{10}$ which can, however, be absorbed by a redefinition of the product of couplings $\Gamma_s\Gamma_b^*$.} as the unprimed ones,
and so we will only consider the two cases ${\cal L}_{{\mathop{\rm int}} }^{a)}$ and ${\cal L}_{{\mathop{\rm int}} }^{b)}$ in the following, including the possible situation of Majorana fermions (neutral scalars).

Through electroweak (EW) symmetry breaking, the SM fermions acquire masses, giving rise to the chirality-flipping process $b\to s\gamma$ and to non-zero contributions to the anomalous magnetic moment of the muon. The corresponding diagrams are shown in Fig.~\ref{fig:radiative}. Note that we do not introduce any additional source of chirality violation beyond the SM. In particular, the Higgs mechanism does not contribute to the masses of the new heavy particles which are supposed to be exclusively generated from explicit mass terms in the respective free-particle Lagrangian. 

Moving from the weak to the mass eigenbasis of the quarks results in a rotation of the couplings $\Gamma^Q_i$ {in flavor space} in Eqs.~(\ref{eq:Linta}) and (\ref{eq:Lintb}). This rotation is unphysical in our setup where we consider the couplings $\Gamma^Q_i$ as independent free parameters. In the mass eigenbasis, we denote the couplings to muons, bottom- and strange-quarks, as $\Gamma_\mu$,  $\Gamma_b$ and $\Gamma_s$, respectively. We further 
assume negligible couplings to the first fermion generation. This assumption allows for an explanation of the 
$R_K$ anomaly and moreover weakens the bounds on the masses of the new particles from direct searches.

Let us now discuss the possible representations for the new particles under the SM gauge group. To this end,
recall that the SM fermions carry the following gauge quantum numbers:
\begin{equation}
\begin{array}{*{20}{c}}
{}&\vline& {SU\!\left(3 \right)}&{SU\!\left( 2 \right)_L}&{U{{\left( 1 \right)}_Y}}\\
\hline
Q&\vline& 3&2&{1/6}\\
u&\vline& 3&1&{2/3}\\
d&\vline& 3&1&{ - 1/3}\\
L&\vline& 1&2&{ - 1/2}\\
e&\vline& 1&1&{ - 1}
\end{array}
\end{equation}
In the case of the non-abelian groups we label the respective representations by their dimension. Applying this notation, the fundamental representations of $SU(3)$ and $SU(2)$ are indicated 
by 3 and 2 in the table above, while the corresponding adjoint representations would be labeled as 8 and 3, and singlets are marked as 1.

In the SM, all particles transform under $SU(2)_L$ and $SU(3)$ either as singlets, in the fundamental, or in the adjoint 
representation. We thus content ourselves with considering these three possibilities also for the 
transformation of the new heavy particles. The requirement of gauge invariance of the Lagrangian
in \eq{eq:Linta} or \eq{eq:Lintb} further acts as a constraint on the allowed combinations of 
representations for the three new particles. We end up with the following possibilities with respect to
$SU(2)$ and $SU(3)$:
\begin{equation}
\begin{array}{*{20}{c}}
\begin{array}{*{20}{c}}
{SU\left( 2 \right)}&\vline& {{\Phi _Q},{\Psi _Q}}&{{\Phi_\ell},{\Psi_\ell}}&{\Psi ,\Phi }\\
\hline
I&\vline& 2&2&1\\
II&\vline& 1&1&2\\
III&\vline& 3&3&2\\
IV&\vline& 2&2&3\\
V&\vline& 3&1&2\\
VI&\vline& 1&3&2\\
\end{array}\quad\quad\quad\quad
\begin{array}{*{20}{c}}
{SU\left( 3 \right)}&\vline& {{\Phi _Q},{\Psi _Q}}&{{\Phi_\ell},{\Psi_\ell}}&{\Psi ,\Phi }\\
\hline
A&\vline& 3&1&1\\
B&\vline& 1&{\bar 3}&3\\
C&\vline& 3&8&8\\
D&\vline& 8&{\bar 3}&3\\
\end{array}
\end{array}
\label{eq:reps}
\end{equation}
The hypercharge $Y$ can be freely chosen for one of the new particles. We define $Y_{\Psi}\equiv X$
for the particle $\Psi$ in model class a) and $Y_\Phi=-X$ for the particle $\Phi$ in model class b).
The values for the other two particles $\Phi_{Q,\ell}$ respectively $\Psi_{Q,\ell}$ are then fixed 
from charge conservation in the Lagrangian (\ref{eq:Linta}) or (\ref{eq:Lintb}):
\begin{equation}
\begin{array}{*{20}{c}}
{Y}&\vline& {{\Phi _Q},{\Psi _Q}}&{{\Phi_\ell},{\Psi_\ell}}&{\Psi ,\Phi }\\
\hline
&\vline& {1/6 + X}&{ - 1/2 + X}&{ - X}
\end{array}
\end{equation}
Motivated by the SM charges, we will assume $X$ to be quantized\footnote{The assumption on the quantization of $X$ has no impact on the phenomenological discussion.} in units of $1/6$ with $-1\le X\le 1$. After EW symmetry breaking, the electric charge $Q_{\text{em}}$ derives from the hypercharge and the third component of $SU(2)$
according to
\begin{equation}
	Q_{\text{em}}=T_3+ Y\,.
\end{equation}

As we {have} found six possibilities (denoted by $I,II,III,IV,V,VI$) for the $SU(2)$ representations, and four possibilities (denoted by $A,B,C,D$) for the $SU(3)$ representations, there are in total 24 scenarios for each model class a) and b). In addition, in each of these scenarios one can freely choose the value of $X$. 

The primed Lagrangian ${\cal L}_{{\mathop{\rm int}} }^{a^\prime)}$ (${\cal L}_{{\mathop{\rm int}} }^{b^\prime)}$) in principle allows for all $SU(2)$ representations, but only representation $I$ and $IV$ can give non-zero contributions to $b\to s\mu^+\mu^-$ processes since the corresponding group factors vanish for the other representations. Concerning $SU(3)$ all options $A,B,C,D$ are permitted (with $\bar{3}\to 3$ for $\Psi_\ell,\Phi_\ell$ in the cases B and D). The hypercharge of $\Psi_\ell,\Phi_\ell$ would change to $1/2-X$.
Therefore, the cases with $SU(2)\in\{I,IV\}$, $SU(3)\in\{A,C\}$ and $X=0$ allow for $\Psi$ ($\Phi$) being a Majorana fermion (a real scalar) contributing to $b\to s\mu^+\mu^-$ and $B_s-\bar{B}_s$ mixing. We will put a special emphasis on this situation in our numerical analysis in Sec.~\ref{sec:numerics} because the presence of additional crossed boxes in $b\to s\ell\ell$ and $B_s-\overline{B}_s$ (see Fig.~\ref{fig:Boxescrossed} and second row in Fig.~\ref{fig:diagrams_Bs-mixing}) can lead to interesting phenomenological consequences. 

\section{Observables and bounds on Wilson Coefficients}
\label{sec:observables}

In the previous section we constructed two classes of NP models aiming at an explanation of the 
$b\to s\mu^+\mu^-$ anomalies through one-loop box contributions. The relevant free model parameters 
governing this transition are the couplings $\Gamma_b$, $\Gamma_s$ and $\Gamma_\mu$ together with the masses
of the three new particles, $m_\Psi$, $m_{\Phi_Q}$, $m_{\Phi_\ell}$ in case a) and $m_\Phi$, $m_{\Psi_Q}$, $m_{\Psi_\ell}$
in case b). Unavoidably, the Lagrangian in \eq{eq:Linta} (in \eq{eq:Lintb}) also generates contributions to
$b\to s\nu\bar\nu$, $b\to s\gamma$ and $B_s-\overline{B}_s$ mixing, and in particular the latter sets an important constraint on the subspace spanned by the couplings $\Gamma_b$, $\Gamma_s$ and the masses $m_\Psi$, $m_{\Phi_Q}$ ($m_\Phi$, $m_{\Psi_Q}$). Furthermore, depending on the coupling $\Gamma_\mu$ and the masses $m_\Psi$, $m_{\Phi_\ell}$ ($m_\Phi$, $m_{\Psi_\ell}$), a contribution to the anomalous magnetic moment $(g-2)_\mu$ of the muon emerges that could have the potential to solve the long-standing anomaly in this observable. A complete phenomenological analysis must take into account all 
these processes. In this section we thus provide the Wilson coefficients needed for their theoretical description 
in the models under consideration and derive the experimental bounds on them. 

\boldmath
\subsection{$b\to s\mu^+\mu^-$}
\label{subsec:bsmumu}
\unboldmath
In our models, the only relevant NP contributions to $b\to s\mu^+\mu^-$ transitions reside inside the effective Hamiltonian
\begin{equation}
 {\cal H}_\mathrm{eff}^{\mu^+\mu^-} =  - \frac{{4{G_F}}}{{\sqrt 2 }}{V_{tb}}V_{ts}^* \left( C_9 \mathcal{O}_9+C_{10} 
 \mathcal{O}_{10} \right)\,,
\end{equation}
with
\bea
 {\mathcal{O}_9}       = \frac{\alpha_\mathrm{EM}}{{4\pi}} \left[\bar s{\gamma^\nu }{P_L}b\right] \left[\bar \mu {\gamma_\nu }\mu\right]  \,,\quad\quad\quad\quad
 {\mathcal{O}_{10}}    = \frac{\alpha_\mathrm{EM}}{{4\pi}}  \left[\bar s{\gamma^\nu
 }{P_L}b\right] \left[\bar \mu {\gamma_\nu }{\gamma^5}\mu \right]\,.
\eea
These operators receive NP contributions from box diagrams, photon- and $Z$-penguins. Since we do not introduce any additional source of $SU(2)$ breaking compared to the SM, $Z$-penguin diagrams are necessarily suppressed by $m_b^2/M_Z^2$ and we will neglect them in the following.

The box contributions are depicted in Fig.~\ref{fig:Boxes} for both {case} $a)$ and $b)$. If the new fermion $\Psi$ (scalar $\Phi$) that couples both to quarks and leptons is in a real representation with respect to all gauge transformations, i.e. if it is a singlet or in the adjoint representation with respect to $SU(2)$ and $SU(3)$ and has hypercharge $X=0$, one can consider the possibility that it is a Majorana fermion (real scalar). In this case, also the crossed diagrams (shown in Fig.~\ref{fig:Boxescrossed}) exist and have to be taken into account. In the models of class a) and b), we have
\begin{equation}
\begin{aligned}
C_9^{{\rm{box}},\, a)} &= - C_{10}^{{\rm{box}},\,  a)} ={\cal N}\frac{{{\Gamma _s}\Gamma _b^*|{\Gamma _\mu }{|^2}}}{{32\pi {\alpha _{{\rm{EM}}}}m_\Psi ^2}}\left( {\chi \eta F\left( {x_Q,x_\ell} \right) + 2{\chi ^{\rm{M}}}{\eta ^{\rm{M}}}G\left( {x_Q,x_\ell} \right)} \right)\,,\\
C_9^{{\rm{box}},\, b)} &= - C_{10}^{{\rm{box}},\, b)} = -{\cal N}\frac{{{\Gamma _s}\Gamma _b^*|{\Gamma _\mu }{|^2}}}{{32\pi {\alpha _{{\rm{EM}}}}m_\Phi ^2}}\left( \chi \eta - {{\chi ^{\rm{M}}}{\eta ^{\rm{M}}}} \right)F\left( {y_Q,y_\ell} \right)\,,
\end{aligned}
\label{eq:C9a}
\end{equation}
with $x_Q=m_{\Phi_Q}^2/m_\Psi^2$, $x_\ell=m_{\Phi_\ell}^2/m_\Psi^2$ and $y_Q=m_{\Psi_Q}^2/m_\Phi^2$, $y_\ell=m_{\Psi_\ell}^2/m_\Phi^2$, respectively. Moreover, we have introduced the abbreviation  
\begin{equation}
	{\cal N}^{-1} = \dfrac{4G_F}{\sqrt{2}}V_{tb} V_{ts}^*\,.
\end{equation}
The dimensionless loop functions are defined as
\begin{equation}
\begin{aligned}
F(x,y) &= 
\frac{1}{(1-x)(1-y)} + \frac{x^2 \log{x}}{(1-x)^2(x-y)} +\frac{y^2  \log{y}}{(1-y)^2(y-x)} \,,\\
G(x,y) &= \frac{1}{(1-x)(1-y)} + \frac{x \log{x}}{(1-x)^2(x-y)} +\frac{y  \log{y}}{(1-y)^2(y-x)} \,,
\end{aligned}
\label{eq:FG}
\end{equation}
and simplify in the limit of equal masses to
\begin{align}
F(1,1) = \frac{1}{3}\,,\qquad G(1,1) = -\frac{1}{6}\,.
\end{align}
The $SU(2)$- and $SU(3)$-factors $\eta$, $\eta^{\text{M}}$ and $\chi$, $\chi^{\text{M}}$  
are tabulated~\footnote{Note that for both the $SU(3)$ and the $SU(2)$ generators we use the canonical normalization $[T^b,T^b]=\delta^{ab}/2$, and that we do not absorb a normalization factor into the couplings
$\Gamma_b,\Gamma_s,\Gamma_\mu$. This convention has to be kept in mind when comparing for instance with SUSY 
results in the literature since Supersymmetry dictates the normalization of the gluino-squark-quark coupling to be $\sqrt{2}g_sT^a_{ij}$.}
in Tab.~\ref{tab:SU2} and Tab.~\ref{tab:SU3}, respectively. 
The term involving the $G$-function in \eq{eq:C9a} stems from the crossed box and is only present if $\Psi$ ($\Phi$) is a Majorana fermion (real scalar). If $\Psi$ ($\Phi$) is a Dirac fermion (complex scalar), $\chi^{\text{M}}$ and $\eta^{\text{M}}$ are zero. We have cross-checked our formulae \eq{eq:C9a} against Ref.~\cite{Bertolini:1990if} 
where results had been given for the gluino-squark and the chargino-squark box in Supersymmetry, corresponding to our representations C-I and A-IV, respectively.
\begin{table}
\begin{equation*}
\begin{array}{*{20}{c}}
{SU\left( 2 \right)}&\vline &\eta& \eta^{\text{M}}=\eta_{B\bar B}^{\text{M}}  &\eta_L& \eta_L^{\text{M}}&\eta_{B\bar B}   & \eta_7 & \widetilde\eta_7 &  \eta_8 &  \eta_{a_\mu} & \widetilde\eta_{a_\mu} &{\eta_{3}} &{\widetilde\eta_{3}}\\
\hline
I                   &\vline & 1    & 1                 & 1    & 1                & 1                                      & -\frac{1}{3}+X		& -X				   & 1   &  -1+X		      &    -X	& 
{1}  & {0}      \\
II                  &\vline & 1    & -                 & 0    & -                & 1                                      & \frac{1}{6} + X		& -\frac{1}{2} - X	   & 1   & -\frac{1}{2} + X	      &  -\frac{1}{2} - X    &{0}     &	{1}   \\
III                 &\vline & 5/16 & -                 & 1/4  & -                & 5/16                                   & -\frac{3}{8} +\frac{3}{4} X  & \frac{1}{8} -\frac{3}{4} X & \frac{3}{4} &  -\frac{7}{8}+\frac{3}{4}X &\frac{1}{8}-\frac{3}{4}X   & {1} &  {-\frac{1}{4}}    \\
IV                  &\vline & 5/16 & 1/16              & 1/16 & 5/16             & 5/16                                   & \frac{1}{4} +\frac{3}{4} X    &  -\frac{1}{2} -\frac{3}{4} X     & \frac{3}{4} & -\frac{1}{4}+\frac{3}{4}X  & -\frac{1}{2}-\frac{3}{4}X  &{-\frac{1}{4}}   &{1} \\
V                   &\vline & 1/4  & -                 & 1/2  & -                & 5/16                                   & -\frac{3}{8} +\frac{3}{4} X & \frac{1}{8} -\frac{3}{4}X & \frac{3}{4}& -\frac{1}{2} + X	      & -\frac{1}{2} -X   &{0}     &	{1}     \\
VI                  &\vline & 1/4  & -                 & 1/2  & -                & 1                                      & \frac{1}{6} + X		&-\frac{1}{2} - X	   & 1   & -\frac{7}{8}+\frac{3}{4}X  & \frac{1}{8}-\frac{3}{4}X  &  {1} &{-\frac{1}{4}} 
\end{array}
\end{equation*}\normalsize
\caption{Table of the $SU(2)$-factors entering the Wilson coefficients for the various processes. Results are given
for the six representations I-VI defined in \eq{eq:reps}.}
\label{tab:SU2}
\end{table}

The photon penguin induces a contribution to $C_9$, whereas it does not generate $C_{10}$ because of the vectorial coupling of the photon to muons. For the cases a) and b), the $C_9$
contribution reads
\begin{equation}
\begin{aligned}
C_9^{\gamma,\, {{a)}}}= {\cal N}\frac{{{\Gamma _s}\Gamma _b^*}}{{2m_\Psi ^2}}\chi_7 \left[ {{\eta _7}{F_9}\!\left( {x_Q} \right) - {{\widetilde \eta }_7}{{G}_9}\!\left( {x_Q} \right)} \right],\;\quad\;
C_9^{\gamma,\, {{b)}}}={\cal N}\frac{{{\Gamma _s}\Gamma _b^*}}{{2m_\Phi ^2}}\chi_7 {\left[ {{\widetilde\eta _7}{\widetilde F_9}\!\left( {y_Q} \right) - {{\eta }_7}{{\widetilde G}_9}\!\left( {y_Q} \right)} \right]},
\end{aligned}
\label{eq:C9ag}
\end{equation}
with
\begin{equation}
\begin{aligned}
F_{9}^{}(x)&=\dfrac{-2x^3+6 \log x+9x^2-18x+11}{36(x-1)^4},&\widetilde F_9(x)&=x^{-1}F_9(x^{-1})\,,\\
G_{9}^{}(x)&=\dfrac{7 -36x+45x^2-16x^3 +6(2x-3)x^2\log x}{36(x-1)^4},&\widetilde G_9(x)&=x^{-1} G_9(x^{-1})\,.
\end{aligned}
\end{equation}
In the simplifying limit of equal masses we have
\bea
F_{9}(1)=-\dfrac{1}{24} \,,\qquad G_{9}^{}(1)=\dfrac{1}{8}\,.
\eea
The terms proportional to ${F_9}$
and ${\widetilde F_9}$ in~ \eq{eq:C9ag} stem from the diagram where the photon is emitted by the scalar $\Phi_{(Q)}$, whereas the terms 
proportional to ${G_9}$ and ${\widetilde G}_9$ stem from the diagram where the photon is emitted by the fermion $\Psi_{(Q)}$. 
The $SU(2)$- and $SU(3)$-factors $\eta_7$, $\widetilde\eta_7$ and $\chi_7$  can again be read off from Tabs.~\ref{tab:SU2} and \ref{tab:SU3}.
In the case where the new scalar and the new fermion are singlets under $SU(2)$, 
$\eta_7$ and  $\widetilde\eta_7$ are 
simply given by the charges of the new particles,
$\widetilde\eta_7=q_\Psi$ and $\eta_7=q_{\Phi_Q}=-1/3-q_\Psi$ for case a),   
$\widetilde\eta_7=q_\Phi$ and $\eta_7=q_{\Psi_Q}=-1/3-q_\Phi$ for case b). 
For higher $SU(2)$ representations, $\eta_7$ and  $\widetilde\eta_7$ 
in addition take care of summing the contributions from each isospin component 
of the new particles. For the representations C-I and A-IV, the results of \eq{eq:C9ag} have again been checked against 
Ref.~\cite{Bertolini:1990if}.

\begin{table}
\begin{equation*}
\begin{array}{*{20}{c}}
{SU\left( 3 \right)}&\vline & \chi=\chi_7 & \chi^{\text{M}} & \chi_{B\bar B} & \chi_{B\bar B}^\text{M} & 
\chi_8& \widetilde\chi_8 & \chi_{a_\mu}{=\chi_Z} & \\
\hline
A&\vline & 1 & 1 & 1& 1 & 1 & 0 & 1 \\
B&\vline & 1 & - & 1& - & 0 & 1 & 3 \\
C&\vline &  4/3 & 4/3 & 11/18 & 1/9  & -1/6 & 3/2  & 8 \\
D&\vline & 4/3 & - & 11/18        & - & 3/2  & -1/6 &  3 \\
\end{array}
\end{equation*}
\caption{Table of the $SU(3)$-factors entering the Wilson coefficients for the various processes. Results are given
for the four representations A-D defined in \eq{eq:reps}.}
\label{tab:SU3}
\end{table}
Unlike the box contribution, the photon penguin does not involve the muon coupling $\Gamma_\mu$ but exclusively depends on the combination $\Gamma_s\Gamma_b^*/m_{\Psi(\Phi)}^2$ constrained from $b\to s\gamma$ and $B_s-\overline{B_s}$ mixing. 
We will explicitly demonstrate in Sec.~\ref{sec:numerics} that the resulting bounds, together with the requirement 
of perturbative couplings $\Gamma_s$ and $\Gamma_b$, 
typically render 
$C_9^{\gamma}$ negligibly small. The same statement applies to the Wilson coefficient $C_7$ of the 
magnetic operator operator $\mathcal{O}_7$ (discussed in Sec.~\ref{sec:bsg}) that contributes to $b\to s\ell^+\ell^-$
transitions in the effective theory via tree-level photon exchange.
Therefore, to a good approximation a solution of the $b\to s\mu^+\mu^-$ anomalies must proceed in our model  
via the pattern $C_9=C_9^{{\rm{box}}}+C_9^{\gamma}\simeq C_9^{{\rm{box}}}\equiv -C_{10}^{{\rm{box}}}=-C_{10}$ 
 and 
$C_7\ll C_9$. The current bounds on the generic scenario $C_9=-C_{10}$, obtained from the combined fit to 
$b\to s \mu^+\mu^-$ data, are~\cite{Descotes-Genon:2015uva,c9bound} 
\bea\label{eq:C9bounds}
-0.81\leq C_9&=&-C_{10}\leq	-0.51\,\quad (\text{at }1\,\sigma)\,,\nonumber\\
-0.97\leq C_9&=&-C_{10}\leq	-0.37\,\quad (\text{at }2\,\sigma)\,,\\
-1.14\leq C_9&=&-C_{10}\leq	-0.23\,\quad (\text{at }3\,\sigma)\,.\nonumber
\eea
These ranges are consistent with the ones determined in Ref.~\cite{Altmannshofer:2014rta}. 

\boldmath
\subsection{$B\to K^{(*)}\nu\bar{\nu}$}
\unboldmath
Following Ref.~\cite{bkvv}, we write the relevant effective Hamiltonian as
\begin{equation}
{{\cal H}_{\rm eff}^{\nu_i\nu_j} } =  - \frac{{4{G_F}}}{{\sqrt 2 }}{V_{tb}}V_{ts}^*\;{{C_L^{ij}}{\mathcal{O}_L^{ij}}}\,,\quad\quad
\text{where}\quad
\mathcal{O}_{L}^{ij} = \frac{\alpha }{{4\pi }} [\bar s{\gamma ^\mu }{P_{L}}b][{{\bar \nu }_i}{\gamma _\mu }\left( {1 - {\gamma ^5}} \right){\nu _j}]\,.
\end{equation}
Due to $SU(2)$ invariance, $b\to s\nu\bar\nu$ is linked to $b\to s\ell^+\ell^-$, implying
\begin{equation}
\begin{aligned}
C_L^{22,\,{a)}} &= {\cal N}\frac{{{\Gamma _s}\Gamma _b^*|{\Gamma _\mu }{|^2}}}{{32\pi {\alpha _{{\rm{EM}}}}m_\Psi ^2}}\left( {\chi {\eta _L}F\left( {x_Q,x_\ell} \right) + 2{\chi ^{\rm{M}}}\eta _L^{\rm{M}}G\left( {x_Q,x_\ell} \right)} \right)\,,\\
C_L^{22,\,b)} &= -{\cal N}\frac{{{\Gamma _s}\Gamma _b^*|{\Gamma _\mu }{|^2}}}{{32\pi {\alpha _{{\rm{EM}}}}m_\Phi ^2}}\left( {{\chi {\eta _L}} - \chi ^{\rm{M}}}\eta _L^{\rm{M}}\right)F\left( {y_Q,y_\ell} \right)\,,
\label{eq:CL22}
\end{aligned}
\end{equation}
with the functions $F$ and $G$ defined in \eq{eq:FG} and $\eta_L$, $\eta_L^{\text{M}}$ and $\chi$, $\chi^{\text{M}}$
given in Tabs.~\ref{tab:SU2} and \ref{tab:SU3}.

Since the different neutrino flavors in the decays $B\to K^{(*)}\nu\bar{\nu}$
are not distinguished experimentally, the total branching ratio, normalized to its SM prediction, reads
\begin{equation}
{R_{K^{(*)}}^{\nu\bar{\nu}}} = 
\dfrac{ \sum\limits_{i,j=1}^3 \left| C_L^{\rm SM}\delta^{ij} + {C_L^{ij}} \right|^2}{3\left| {C_L^{\rm SM}} \right|^2} \,,
\end{equation}
where $C_L^{\rm SM}\approx-1.47/\sin^2\theta_w=-6.35$ with $\theta_w$ being the weak mixing angle.
The current experimental limits for $B\to K^{(*)}\nu\bar{\nu}$ are~\cite{bkvv}  (at $90\,\%$ C.L.)
\begin{equation}
{R_K^{\nu\bar{\nu}}} < 4.3	\,,\qquad {R_{{K^*}}^{\nu\bar{\nu}}} <4.4\,.
\end{equation}
While $C_L^{22}$, given in \eq{eq:CL22}, involves the muonic coupling $\Gamma_\mu$, any other coefficient $C_L^{ij}$ 
with $(i,j)\neq(2,2)$ would depend on the couplings $\Gamma_e$, $\Gamma_\tau$ of the new particles to electrons or tauons. 
Since we do not want to make any assumptions on the size of these couplings, we will implement the bound from $B\to K^{(*)}\nu\bar{\nu}$ according to
\bea
\left| 1+  \dfrac{C_L^{22}}{C_L^{\rm SM}} \right|^2 \le 
\sum\limits_{i,j=1}^3 \left| \delta^{ij}+  \dfrac{C_L^{ij}}{C_L^{\rm SM}} \right|^2\le 3{R_{{K^{(*)}}}^{\nu\bar{\nu}}}
\le12.9\quad \text{(at  90\% C.L.)},
\label{eq:clbound}
\eea
leading to the following bound on $C_L^{22}$:
\begin{equation}
   -16.5\le C_L^{22} \le 29.2  \quad \text{(at  90\% C.L.)}.
\end{equation}
Since this constraint is more than an order of magnitude weaker 
than the bound in \eq{eq:C9bounds} on the $SU(2)$-related coefficient $C_9$ of $b\to s\mu^+\mu^-$, we will not consider it in our numerical analysis.

\boldmath
\subsection{$B_s-\overline{B}_s$ mixing}
\label{suse:BsMix}
\unboldmath
\begin{figure*}[thb]
\begin{center}
\begin{tabular}{cp{7mm}c}
\includegraphics[width=0.9\textwidth]{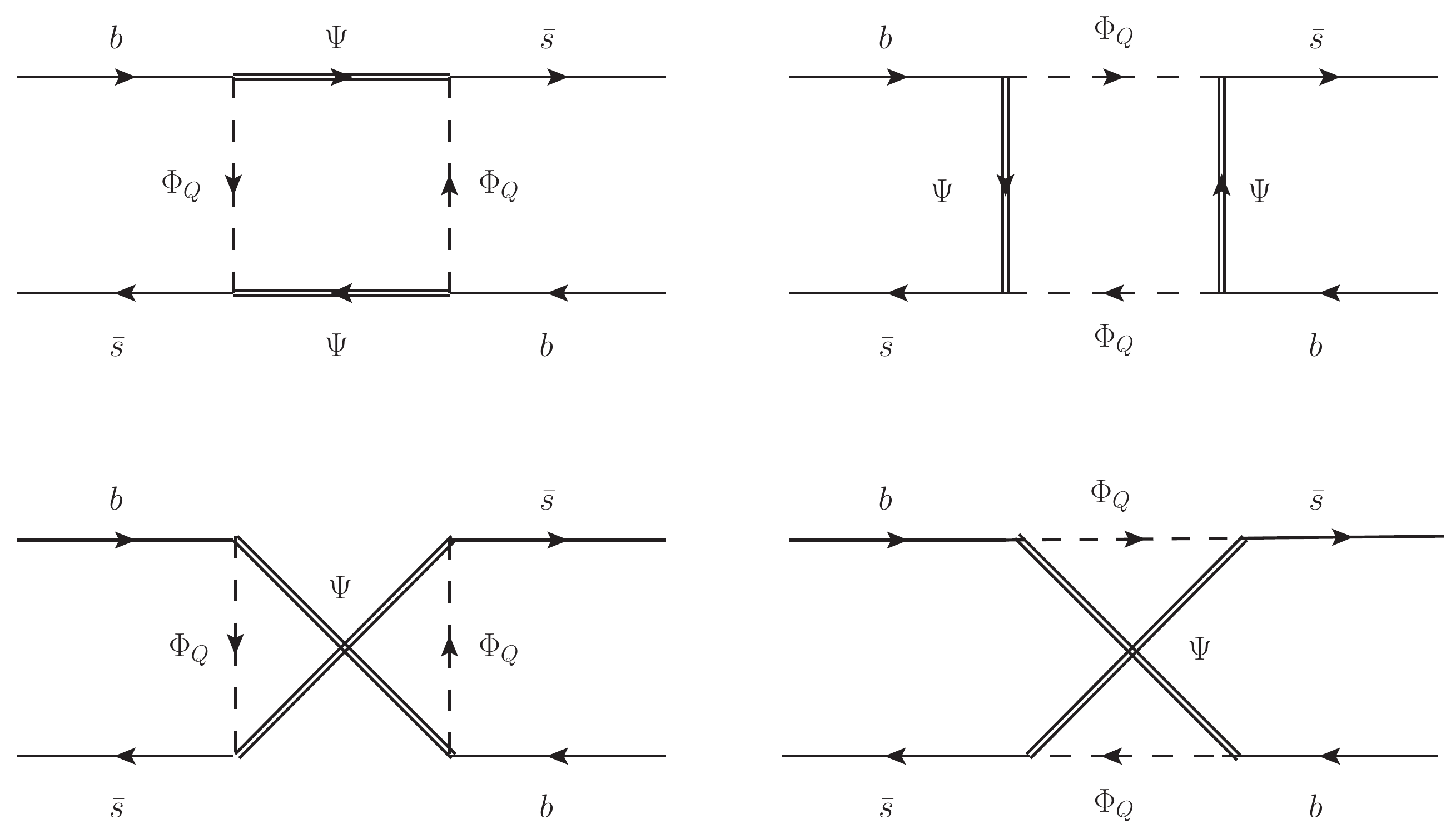}
\end{tabular}
\end{center}
\caption{Loop contributions to $B_s-\overline{B}_s$ mixing in the case $a)$. The crossed diagrams only exist if $\Psi$ is
a Majorana fermion. The case $b)$ is obtained by the replacement $\Psi\to \Phi_Q$ and $\Phi_Q\to \Psi$.
 \label{fig:diagrams_Bs-mixing} }
\end{figure*}

Contributions to $B_s-\overline{B}_s$ mixing arise from box diagrams mediated in models of class a) by the fermion $\Psi$ and the scalar $\Phi_Q$ and in models of class b) by the fermion $\Psi_Q$ and the scalar $\Phi$ (see Fig.~\ref{fig:diagrams_Bs-mixing}). If $\Psi$ is a Majorana fermion (or $\Phi$ is a real scalar), also the corresponding crossed boxes have to be taken into account. Since the new particles only couple to left-handed SM fermions, the effective Hamiltonian only involves one operator:
\begin{equation}
{\cal H}_{\rm eff}^{B\bar B}=
 C_{B\bar B}
 ({\bar s}_{\alpha} \gamma^{\mu} P_{L} b_{\alpha})\, ({\bar s}_{\beta} \gamma^{\mu} P_{L} b_{\beta})     \,,\\
\label{opbasisDF2}
\end{equation}
where $\alpha$ and $\beta$ are color indices. The NP contribution to the Wilson coefficient reads
\begin{equation}
\begin{aligned}
C_{B\bar B}^{a)} &= \frac{{{{\left( {{\Gamma _s}\Gamma _b^*} \right)}^2}}}{{128{\pi ^2}m_\Psi ^2}}\left( {{\chi _{B\bar B}}{\eta _{B\bar B}}F\left( {x_Q,x_Q} \right) + 2\chi _{B\bar B}^{\rm{M}}\eta _{B\bar B}^{\rm{M}}G\left( x_Q,x_Q \right)} \right)
\,,\\
C_{B\bar B}^{b)} &= \frac{{{{\left( {{\Gamma _s}\Gamma _b^*} \right)}^2}}}{{128{\pi ^2}m_\Phi ^2}}\left( {{\chi _{B\bar B}}{\eta _{B\bar B}} - \chi _{B\bar B}^{\rm{M}}\eta _{B\bar B}^{\rm{M}}} \right)F\left( {y_Q,y_Q} \right)\,,
\end{aligned}
\label{eq:Cbb}
\end{equation}
with the loop functions $F$ and $G$ defined in \eq{eq:FG} and $\eta_{B\bar B}$, $\eta_{B\bar B}^{\text{M}}$ and  $\chi_{B\bar B}$, $\chi_{B\bar B}^{\text{M}}$ given in Tabs.~\ref{tab:SU2} and \ref{tab:SU3}.
For the representations C-I and A-IV, \eq{eq:Cbb} agrees with 
the results of Ref.\cite{dms-check} for the gluino-squark and the chargino-squark boxes.

To derive bounds on the Wilson coefficient $C_{B\bar B}(\mu_H)$, we define the ratio
\begin{equation}
{R_{\Delta B_s}} = \dfrac{\Delta M_{B_s}^{\rm exp}}{\Delta M_{B_s}^{\rm{SM}}}-1=
\dfrac{C_{B\bar B}(\mu_H)}{C_{B\bar B}^{\rm SM}(\mu_H)} \,.
\end{equation}
For the SM prediction $\Delta M_{B_s}^{\rm{SM}}$, we use $C_{B\bar B}^{\rm SM}(\mu_H)\simeq 8.2 \times 10^{-5}{\rm TeV}^{-2}$ at $\mu_H=2 m_W$, together with the recent lattice results of Ref.~\cite{Bazavov:2016nty} for the hadronic matrix element $f_{B_s}^2 B_{B_s}$. With this input, we find $R_{\Delta B_s}=-0.09\pm 0.08$, i.e. the experimental value $\Delta M_{B_s}^{\rm{exp}}$ is below the SM prediction $\Delta M_{B_s}^{\rm{SM}}$ by about $1\sigma$. Note that the lattice results of Ref.~\cite{Bazavov:2016nty} have not yet been  included in the 2016  FLAG report~\cite{Aoki:2016frl}. They are compatible with the FLAG average~\cite{Aoki:2016frl} but are supposed to be more precise by roughly a factor two. As an updated average of  $f_{B_s}^2 B_{B_s}$ from FLAG, including also~\cite{Bazavov:2016nty}, is not yet available, we directly use Ref.~\cite{Bazavov:2016nty} for our numerical analysis. For the bound on $C_{B\bar B}(\mu_H)$  we finally get~\footnote{By using the 2016 FLAG  average~\cite{Aoki:2016frl} we would get $C_{B\bar B}(\mu_H) \in [-2.3,1.6]\times 10^{-5}\, \rm TeV^{-2}\:\: (\text{at } 2\, \sigma)$.}
\begin{equation}
\begin{aligned}
C_{B\bar B}(\mu_H) &\in& [-2.1,0.6]\times 10^{-5}\, \rm TeV^{-2}\quad (\text{at } 2\, \sigma),\\
C_{B\bar B}(\mu_H) &\in& [-2.8,1.3]\times10^{-5}\, \rm TeV^{-2}\quad (\text{at } 3\, \sigma)\,.
\label{eq:boundCBB}
\end{aligned}
\end{equation}

Note that the $SU(2)_L$ symmetry of the SM links the up-type couplings $\Gamma_u$ to the down-type couplings through a CKM rotation. Therefore, non-vanishing couplings to up-quarks are generated in our model, namely 
\begin{equation}
  \Gamma_u\,=\,V_{us}\Gamma_s+V_{ub}\Gamma_b\hspace{0.5cm}\textrm{and}\hspace{0.5cm}\Gamma_c\,=\,V_{cs}\Gamma_s+V_{cb}\Gamma_b\,.
\label{Gammau}
\end{equation}
 These couplings control the size of the contributions to $D_0-\bar{D}_0$ mixing. 
The corresponding  coefficient $C_{D\bar D}$  is obtained from $C_{B\bar B}$ by replacing
$\Gamma_s\to \Gamma_u$ and $\Gamma_b\to \Gamma_c$ in Eq.~(\ref{eq:Cbb}).

Since a precise SM prediction for $D_0-\bar{D}_0$ is lacking, we constrain the NP contribution to $C_{D\bar D}$ by the requirement that it does not generate a larger mass difference 
than the one measured experimentally:
\begin{equation}
\begin{aligned}
|C_{D\bar D}(\mu_H)| &<& 2.7\times 10^{-7}\, \rm TeV^{-2}\quad (\text{at } 2\, \sigma),\\
|C_{D\bar D}(\mu_H)| &<& 3.4\times10^{-7}\, \rm TeV^{-2}\quad (\text{at } 3\, \sigma)\,.\label{eq:boundCDD}
\end{aligned}
\end{equation}
To obtain these bounds, we used the recent results for the $D_0-\bar{D}_0$ system in Ref.~\cite{luca} and lattice inputs from Ref.~\cite{nuria}.

\begin{figure*}[t]
\begin{center}
\begin{tabular}{cp{7mm}c}
\includegraphics[width=0.9\textwidth]{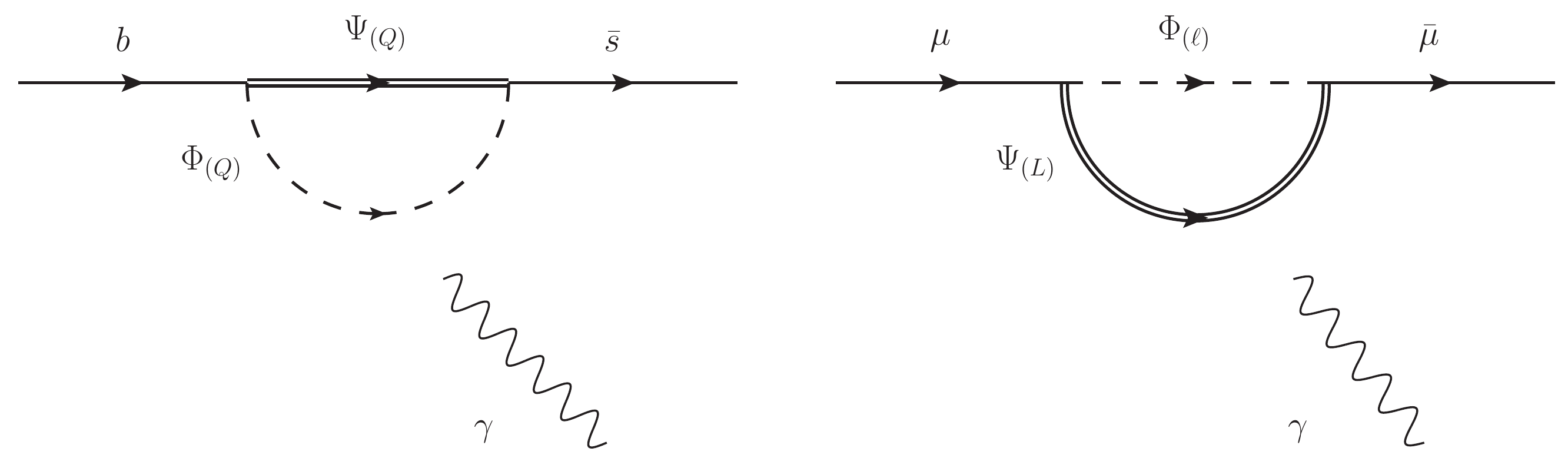}
\end{tabular}
\end{center}
\caption{Loop contributions to $b\to s\gamma$ and the anomalous magnetic moment of the muon. \label{fig:radiative} }
\end{figure*}

\boldmath
\subsection{$b\to s\gamma$}
\label{sec:bsg}
\unboldmath

In our models, $b\to s\gamma$ is affected by NP contributions to the effective Hamiltonian 
\begin{equation}
  {\cal H}_{\rm eff}^{\gamma}=-\frac{4G_F}{\sqrt{2}}V_{tb}V^*_{ts} (C_7 \mathcal{O}_7+C_8\mathcal{O}_8)\,,
\end{equation}
with
\begin{eqnarray}
\mathcal{O}_7=\frac{e}{16\pi^2} m_b \bar{s}\sigma^{\mu\nu} P_R b F_{\mu \nu}\,,\quad\quad
\mathcal{O}_8=\frac{g_s}{16\pi^2} m_b \bar{s}_\alpha\sigma^{\mu\nu} P_R T^a_{\alpha\beta}b_\beta G_{\mu \nu}^a\,.
\end{eqnarray}
Here, $F_{\mu\nu}$ and $G_{\mu\nu}^a$ are the field strength {tensors} of the photon and the gluon field, respectively.
While the operator $\mathcal{O}_7$ generates the process $b\to s\gamma$ at tree-level, the operator $\mathcal{O}_8$ contributes via its QCD mixing into $\mathcal{O}_7$.

In the cases a) and b) we find the Wilson coefficients
\begin{equation}
C_7^{a)}={\cal N}\frac{{{\Gamma _s}\Gamma _b^*}}{{2m_\Psi ^2}}{\chi _7}\left[ {{\eta _7}F_7^{}\left( {x_Q} \right) - {{\widetilde \eta }_7}\widetilde F_7^{}\left( {x_Q} \right)} \right],\quad\quad
C_8^{a)}={\cal N}\frac{{{\Gamma _s}\Gamma _b^*}}{{2m_\Psi ^2}}{\eta _8}\left[ {{\chi _8}F_7^{}\left( {x_Q} \right) - {{\widetilde \chi }_8}\widetilde F_7^{}\left( {x_Q} \right)} \right],
\label{eq:C7a}
\end{equation}
and
\begin{equation}
C_7^{b)}={\cal N}\frac{{{\Gamma _s}\Gamma _b^*}}{{2m_\Phi ^2}}{\chi _7}\left[ {{\widetilde\eta _7}\widetilde F_7^{}\left( {y_Q} \right) - {{ \eta }_7} F_7^{}\left( {y_Q} \right)} \right],\quad\quad
C_8^{b)}={\cal N}\frac{{{\Gamma _s}\Gamma _b^*}}{{2m_\Phi ^2}}{\eta _8}\left[ {{\widetilde\chi _8}\widetilde F_7^{}\left( {y_Q} \right) - {{\chi }_8} F_7^{}\left( {y_Q} \right)} \right].
\label{eq:C7b}\end{equation}
The loop functions are given by
\begin{equation}
F_7^{}(x)=\frac{ x^3-6 x^2+6 x \log x+3 x+2}{12 (x-1)^4}\,,\quad\quad
\widetilde F_7^{}(x)=x^{-1} F_{7}(x^{-1})\,,
\end{equation}
taking the value $F_7^{}(1)=\widetilde F_7^{}(1)=1/24$ in the limit of equal masses. For the $SU(2)$- and $SU(3)$-factors  $\eta_7,\widetilde\eta_7,\eta_8$ and $\chi_7,\chi_8,\widetilde\chi_8$ 
we refer the reader to Tabs.~\ref{tab:SU2} and \ref{tab:SU3} as usual.
As in the case of $C_9^\gamma$, 
we identify $\widetilde\eta_7,\eta_7$ with the charges of the new particles if they are $SU(2)$ singlets.
Our results of $C_{7,8}$ for the representations C-I and A-IV are in agreement 
with the ones of Refs.\cite{bsg-check1,bsg-check2} for the gluino-squark and the chargino-squark contributions 
in Supersymmetry.
	
The most recent experimental result~\cite{Amhis:2014hma} and SM prediction~\cite{Misiak:2015xwa} for the branching ratio of $b\to s\gamma$ are given by
\begin{eqnarray*}
{\rm Br}^{\rm exp}[b\to s\gamma]&=&(3.43 \pm 0.21 \pm 0.07) \times 10^{-4}\,,\\
{\rm Br}^{\rm SM}[b\to s\gamma]&=&(3.36 \pm 0.23) \times 10^{-4}\,.
\end{eqnarray*}
In order to implement the constraint from $b\to s\gamma$ on the NP coefficients $C_7,C_8$
(defined at the high scale $\mu_H=2 m_W)$, we introduce 
the ratio~\footnote{$C_{7,8}$ in Eqs.~(\ref{eq:C7a},\ref{eq:C7b}) are given in the same sign convention as  
$C_{7,8}^{\rm{SM}}$ in Ref.~\cite{Misiak:2015xwa}, where 
$C_{7}^{\rm{SM}}(\mu_H)=-0.197$ and $C_{8}^{\rm{SM}}(\mu_H)=-0.098$ at leading order in QCD.}
\bea
R_{b\to s\gamma} &=& \dfrac{\text{Br}^{\rm exp}[b\to s\gamma]}{\text{Br}^{\rm SM}[b\to s\gamma]}-1= 
- 2.45\, \left[C_7(\mu_H)+0.24\,C_8(\mu_H)\right]
\eea
where the combination $C_7+0.24\, C_8$ takes into account QCD effects~\cite{Misiak:2015xwa}. Adding the statistical and the systematic experimental error in quadrature, and combining it linearly with the theory error linearly, we find 
$-0.17\leq R_{b\to s\gamma}\leq 0.24$ at the $2\,\sigma$ level, being equivalent to 
\bea
-0.098\leq C_7(\mu_H)+0.24\,C_8(\mu_H)\leq	0.070\,\quad (2\,\sigma)\,.
\eea
 
\boldmath 
\subsection{Anomalous magnetic moment of the muon}
\unboldmath

The anomalous magnetic moment (AMM) of the muon, $a_\mu \equiv (g-2)_\mu/2$, also receives a NP contribution 
in our setup. 
Using the effective Hamiltonian (see for example~\cite{Crivellin:2012jv})
\begin{equation}
{\cal H}_{\rm eff}^{a_\mu}={-}a_\mu \dfrac{e}{4 m_\mu} 
 (\bar{\mu}\sigma^{\mu\nu} \mu)\, F_{\mu \nu},
\end{equation}
we find
\begin{equation}
\begin{aligned}
\Delta a_\mu^{{a)}}&=\frac{m_\mu^2 |\Gamma_\mu|^2}{8\pi^2m_\Psi^2}
\chi_{a_\mu}\left[\eta_{a_\mu} F_{7}(x_\ell)-\widetilde\eta_{a_\mu} \widetilde F_{7}(x_\ell)\right],\\
\Delta a_\mu^{{b)}}&=\frac{m_\mu^2 |\Gamma_\mu|^2}{8\pi^2m_\Phi^2}
\chi_{a_\mu}\left[\widetilde\eta_{a_\mu}\widetilde F_{7}(y_\ell)-\eta_{a_\mu}  F_{7}(y_\ell)\right].
\end{aligned}
\label{eq:amm}
\end{equation}
The group factors  $\eta_{a_\mu}$, $\widetilde\eta_{a_\mu}$ and $\chi_{a_\mu}$ are again given
in Tabs.~\ref{tab:SU2} and \ref{tab:SU3}. If the new particles are SU(2) singlets, 
we have $\widetilde\eta_{a_\mu}=q_\Psi$ and $\eta_{a_\mu}=q_{\Phi_\ell}=-1-q_{\Psi}$ for case a),   
and $\widetilde\eta_{a_\mu}=q_\Phi$ and $\eta_{a_\mu}=q_{\Psi_\ell}=-1-q_\Phi$ for case b).
Our result for $\Delta a_\mu$  has 
been cross-checked for the representation A-IV by comparison with the chargino-squark 
and the neutralino-squark results in Refs.~\cite{gm2-check1,Moroi:1995yh}.
	
The experimental value of $a_\mu^\mathrm{exp} = (116\,592\,091\pm54\pm33) \times 10^{-11}$ (where the first error is statistical and the second systematic) is completely dominated by the Brookhaven experiment E821~\cite{Bennett2006}. The SM prediction is given by~\cite{Aoyama:2012wk,Czarnecki:1995sz,Czarnecki:1995wq,Gnendiger:2013pva,Davier:2011zz,Hagiwara:2011af,Kurz:2014wya,Jegerlehner:2009ry,Colangelo2014} $a_\mu^\mathrm{SM} = (116\,591\,855\pm59) \times 10^{-11}$, where almost the entire uncertainty is due to hadronic effects. The difference between the SM prediction and the experimental value,
\begin{equation} \label{eq:g-2}
 \Delta a_\mu = a_\mu^\mathrm{exp}-a_\mu^\mathrm{SM} = (236\pm 87)\times 10^{-11},
\end{equation}
amounts to a $2.7\sigma$ deviation\footnote{Less conservative estimates lead to discrepancies up to $3.6\,\sigma$}.

The measurement of $R_K$ by LHCb hints towards lepton-flavour universality violation. In global fits to the full set of $b\to s\ell^+\ell^-$ data this manifests itself
as a preference for scenarios with NP contributions $|C_9^e|\ll |C_9^\mu|$~\cite{Altmannshofer:2014rta,Descotes-Genon:2015uva}. In our model this pattern transforms into 
$|\Gamma_e|\ll|\Gamma_\mu|$, and for simplicity we assume $\Gamma_e=0$ in our phenomenological analysis. 
In the presence of a non-zero $\Gamma_e$, the transition $\mu\to e\gamma$ is generated in a similar manner as $a_\mu$ and the measured branching ratio sets a constraint
on the product $\Gamma_\mu\Gamma_e^*$.

The decay $\mu\to e\gamma$ is described by the effective Hamiltonian
\begin{equation}
{\cal H}_{\rm eff}^{\mu \to e\gamma}={-}C_{\mu \to e\gamma}  m_\mu 
 (\bar{e}\sigma^{\mu\nu} P_R \mu)\, F_{\mu \nu},
\end{equation}
from which the branching ratio is obtained according to
\bea
\textrm{Br}\left( {\mu  \to e\gamma } \right) = \frac{{m_\mu ^5}}{{4\pi }}{\tau _\mu }
|C_{\mu \to e\gamma}|^2,
\eea
where $\tau_\mu$ denotes the life-time of the muon. In our models, the Wilson coefficient $C_{\mu \to e\gamma}$ is directly related 
to the NP contribution to the anomalous magnetic moment of the muon as
\begin{equation}
  C_{\mu \to e\gamma}\,=\,\frac{e}{m_\mu^2}\frac{\Gamma_e^*}{\Gamma_\mu^*}\Delta a_\mu.
  \label{eq:relemu}
\end{equation}

The experimental upper limit~\cite{TheMEG:2016wtm} is currently given by
\begin{equation}
 \textrm{Br}^{\textrm{exp}}\left( {\mu  \to e\gamma } \right) \leq 4.2\times 10^{-13},
\end{equation}
which translates into
\begin{equation}
   m_\mu^2|C_{\mu \to e\gamma}|\,<\,3.9\times 10^{-15}
\end{equation}
for the Wilson coefficient. The relation \eq{eq:relemu} between $a_\mu$ and $\mu\to e\gamma$ then implies that a solution of anomaly in $a_\mu$ requires
a strong suppression of $\Gamma_e$ with respect to $\Gamma_\mu$. Already a minimal shift $\Delta a_\mu=61\times 10^{-11}$, as needed to reduce the tension
from $2.7\sigma$ to $2.0\sigma$, is consistent with the bound from $\mu\to e\gamma$ only for $|\Gamma_e/\Gamma_\mu|<2\times 10^{-5}$.

\boldmath
\subsection{$Z\mu^+\mu^-$ coupling}
\label{sec:Zmumu}
\unboldmath
Exchanging the photon in the diagrams of Fig.~\ref{fig:radiative} with the $Z$ boson, effective $Zq_i\bar{q}_j$ and $Z\mu^+\mu^-$ vertices are generated.
Note that our model does not break the $SU(2)_L$ symmetry of the SM and that the $Z$ boson acts like a $U(1)_Z$ gauge boson 
in neutral-current processes in the absence of $SU(2)_L$-breaking sources. For this reason the QED Ward identity holds for the NP corrections to the $Zq_i\bar{q}_j$ and $Z\mu^+\mu^-$ vertices and it follows that the vertex correction and the fermionic field renormalization for on-shell fermions 
cancel in the limit $q^2\to 0$ with $q$ being the momentum carried by the (off-shell) $Z$ boson\footnote{The correction to the 
self-energy of the $Z$ boson does not cancel but involves the weak gauge coupling and not the potentially large new couplings $\Gamma_{b,s,\mu}$.}. 
This implies that the NP contribution exhibits a $q^2/m_{\Psi(\Phi)}^2$ suppression when the vertex is probed for $q^2\ll m_{\Psi(\Phi)}^2$, rendering
the $Z$-penguin contribution irrelevant for $B$ decays where $q^2=\mathcal{O}(m_b^2)$. At LEP, however, the couplings of the $Z$ boson
have been measured for $q^2=M_Z^2$ and the less severe suppression of the NP contribution at this scale together with the high precision of the LEP 
data could lead to relevant constraints for the model.

The LEP bounds are most important for the $Z\mu^+\mu^-$ coupling because this coupling has been determined most accurately and, moreover, 
the corrections involve the coupling $\Gamma_\mu$ which is required to be large to solve both the $b\to s\mu^+\mu^-$ and the $a_\mu$ anomalies. As mentioned above, 
the $Z$ boson behaves like a heavy photon in the $Z$ penguin contribution and the corresponding formula is thus related to the 
one of the photon penguin in Eq.~(\ref{eq:C9ag}). The correction proportional to $|\Gamma_\mu|^2$ to the left-handed $Z\mu^+\mu^-$ coupling
is given by
\begin{equation}
\begin{aligned}
&\frac{\delta g_{L\,\mu}^{a)}}{g_{L\,\mu}^{\textrm{SM}}}(q^2)\;=\;\frac{1}{32\pi^2}\left(\frac{1}{1-2s_W^2}\right)\frac{q^2}{m_\Psi^2}
      |\Gamma _\mu|^2\chi_Z \left[ {{\eta _Z}{F_9}\!\left( {x_\ell} \right) - {{\widetilde \eta }_Z}{{G}_9}\!\left( {x_\ell} \right)} \right],\;\quad\;\\
&\frac{\delta g_{L\,\mu}^{b)}}{g_{L\,\mu}^{\textrm{SM}}}(q^2)\;=\;\frac{1}{32\pi^2}\left(\frac{1}{1-2s_W^2}\right)\frac{q^2}{m_\Psi^2}
      |\Gamma _\mu|^2\chi_Z {\left[ {{\widetilde\eta _Z}{\widetilde F_9}\!\left( {y_\ell} \right) - {{\eta }_Z}{{\widetilde G}_9}\!\left( {y_\ell} \right)} \right]},
\end{aligned}
\label{eq:Zmumu}
\end{equation}
where 
$\eta_Z= \eta_3 +2s_W^2\eta _{a_\mu }$ and $\widetilde\eta_Z=\widetilde\eta _3+2s_W^2\widetilde\eta _{{a_\mu }}$.
The group factors $\chi_Z$, $\eta_3$, $\widetilde{\eta}_3$ are again given in Tabs.~\ref{tab:SU2} and \ref{tab:SU3}, and we have introduced
the abbreviation $s_W=\sin\theta_W$ with $\theta_W$ being the weak mixing angle. For the representation A.II in case a) 
(A.I in case b)), our model generates the same NP contribution to the $Z\mu^+\mu^-$ coupling as the model considered in Ref.~\cite{Belanger:2015nma}, 
and we explicitly cross-checked our formulae \eq{eq:Zmumu} for this special case against the corresponding formula in \cite{Belanger:2015nma}. 

From the LEP measurement~\cite{ALEPH:2005ab} $g_{L\,\mu}^{\textrm{exp}}(m_Z^2)=-0.2689\pm 0.0011$ we infer the following bound at the 2$\sigma$ level:
\begin{equation}
  \left|\frac{\delta g_{L\,\mu}}{g_{L\,\mu}^{\textrm{SM}}}(m_Z^2)\right| \leq 0.8\%\;\;\;(2\sigma).
  \label{eq:Zmumu_LEP}
\end{equation}

\section{Phenomenological analysis}
\label{sec:numerics}

The processes described in the previous section depend in our models on five independent free parameters: the product of couplings $\Gamma_s^*\Gamma_b$ and the absolute value of the coupling $|\Gamma_\mu|$, as well as the three masses $m_{\Psi(\Phi)}$, $m_{\Phi_Q(\Psi_Q)}$, $m_{\Phi_\ell(\Psi_\ell)}$. The decay $b\to s\gamma$ and $B_s-\overline{B}_s$ mixing, both exclusively related to the quark sector, are experimentally and theoretically very precise observables and thus set stringent constraints on the subspace spanned by $\Gamma_s^*\Gamma_b$ and $m_{\Psi(\Phi)}$, $m_{\Phi_Q(\Psi_Q)}$. In this section we will address the question whether these constraints still allow to choose $|\Gamma_\mu|$ and $m_{\Phi_\ell(\Psi_\ell)}$ in such way that a solution of the anomalies in $b\to s\mu^+\mu^-$ and $a_\mu$ is {provided}. 

Since the loop functions that appear in the Wilson coefficients are smooth functions of the squared mass ratios,
the general phenomenological features can {in a first approximation} be studied in the limit of equal masses $m_{\Psi(\Phi)}=m_{\Phi_Q(\Psi_Q)}=m_{\Phi_\ell(\Psi_\ell)}$, reducing the number of free parameters from five to three. The corresponding analysis will be presented in Sec.~\ref{sec:eqmass}. An exception occurs if $\Psi$ is a Majorana fermion: In this case we encounter negative interference {between} the loop functions $F$ and $G$ in the coefficient $C_{B\overline{B}}$ which can be used to avoid or to weaken the stringent bound from $B_s-\overline{B}_s$ mixing in a setup with unequal masses of the new particles. This possibility will be discussed in Sec.~\ref{sec:uneqmass}.
 
\begin{table*}
\[
\begin{array}{*{20}{c}}
{\xi_{B\bar B}} & \vline& I&{II}&{III}&{IV}&V&{VI}\\
\hline
A&\vline& {1\;\left( 0 \right)}&{1}&{\frac{5}{{16}}}&{\frac{5}{{16}}\;\left( {\frac{1}{{4}}} \right)}&{\frac{5}{{16}}}&{1}\\
B&\vline& {1}&{1}&{\frac{5}{{16}}}&{\frac{5}{{16}}}&{\frac{5}{{16}}}&{1}\\
C&\vline& {\frac{{11}}{{18}}\;\left( {\frac{1}{2}} \right)}&{\frac{{11}}{{18}}}&{\frac{{55}}{{288}}}&{\frac{{55}}{{288}}\;\left( {\frac{{53}}{{288}}} \right)}&{\frac{{55}}{{288}}}&{\frac{{11}}{{18}}}\\
D&\vline& {\frac{{11}}{{18}}}&{\frac{{11}}{{18}}}&{\frac{{55}}{{288}}}&{\frac{{55}}{{288}}}&{\frac{{55}}{{288}}}&{\frac{{11}}{{18}}}
\end{array}\hspace{2cm}
\begin{array}{*{20}{c}}
{\xi_9^{{\rm{box}}}}&\vline& I&{II}&{III}&{IV}&V&{VI}\\
\hline
A&\vline& {1\;\left( 0 \right)}&1&{\frac{5}{{16}}}&{\frac{5}{{16}}\;\left( {\frac{1}{{4}}} \right)}&{\frac{1}{{4}}}&{\frac{1}{{4}}}\\
B&\vline& {1}&{1}&{\frac{5}{{16}}}&{\frac{5}{{16}}}&{\frac{1}{{4}}}&{\frac{1}{{4}}}\\
C&\vline& {\frac{4}{3}\;\left( 0 \right)}&{\frac{4}{3}}&{\frac{5}{{12}}}&{\frac{5}{{12}}\;\left( {\frac{1}{3}} \right)}&{\frac{1}{3}}&{\frac{1}{3}}\\
D&\vline& {\frac{4}{3}}&{\frac{4}{3}}&{\frac{5}{{12}}}&{\frac{5}{{12}}}&{\frac{1}{3}}&{\frac{1}{3}}
\end{array}
\]
\caption{Group factor for $B_s-\overline{B}_s$ mixing and $C_9^{\rm box}$ for the case equal masses. The number in brackets are for the case of Majorana fermions or real scalars. }
\label{CBBC9factors}
\end{table*}

\boldmath
\subsection{Degenerate masses: $m_{\Psi(\Phi)}=m_{\Phi_Q(\Psi_Q)}=m_{\Phi_\ell(\Psi_\ell)}$}
\label{sec:eqmass}
\unboldmath

Under the assumption of equal masses $m_{\Psi(\Phi)}=m_{\Phi_Q(\Psi_Q)}=m_{\Phi_\ell(\Psi_\ell)}$, both setups a) 
and b) give identical results for all Wilson coefficients and we can discuss them together. We will denote the common mass as $m_\Psi$ in the following. As a benchmark point we will assume a mass of 1 TeV which is save with respect to direct LHC searches from Run I and current Run II data. The collider signature of our model is similar to the one of sbottom searches in the MSSM if the fermion is not charged under QCD and electrically neutral. The corresponding mass limits at the LHC with 13 TeV can reach up to 800 GeV from Atlas and CMS~\cite{Aaboud:2016nwl,Khachatryan:2016kod}. Note further that the limits 
strongly depend on the embedding of the set-up in a more complete theory and that the bounds can be expected to be significantly 
weaker in our case since we assume approximately degenerate $m_{\Psi(\Phi)}\approx m_{\Phi_Q(\Psi_Q)}\approx
m_{\Phi_\ell(\Psi_\ell)}$\footnote{The non-degenerate case can actually give a rich phenomenology still allowing mass limits well below 1 TeV  (see Ref.~\cite{Gripaios:2015gra} for an analysis of Run I data). 
We will consider LHC limits in more detail in a future work including final Run II results.}.

It turns out that $B_s-\overline{B}_s$ mixing imposes very stringent constraints in the $(\Gamma_s^*\Gamma_b,m_\Psi)$-plane.
This is caused by the fact that $C_{B\bar B}$ is positive and thus increases $\Delta M_{B_s}$, pushing it even further away from the experimental {central} value. At $2\sigma$, we find
\begin{equation}
|\Gamma_s^*\Gamma_b|\le \:0.15\:\frac{1}{\sqrt{\xi_{B\bar B}}}\;\frac{m_\Psi}{1\,{\rm TeV}}\,, 
\end{equation}
where the combinatorial factor $\xi_{B\bar B}={{\chi _{B\bar B}}{\eta _{B\bar B}} - \chi _{B\bar B}^{\rm{M}}\eta _{B\bar B}^{\rm{M}}}$, tabulated in Tab.~\ref{CBBC9factors}, can weaken the bound at most by a factor $1/\sqrt{(\xi_{B\bar B})_{\rm min}}\approx 2.3$. The constraint heavily affects the photon penguin contributions to $C_7$ and $C_9^\gamma$  which depend on the same free parameters $\Gamma_s^*\Gamma_b$ and $m_\Psi$. In the most favorable representation, 
and allowing for hypercharges $X\in[-1,+1]$, we find these contributions to be completely negligible:
\begin{equation}
   |C_7+0.24C_8|\;\le\; 0.018\;\frac{1\,{\rm TeV}}{m_\Psi}\,,\hspace{1cm}
   |C_9^\gamma|\;\le\; 0.02\;\frac{1\,{\rm TeV}}{m_\Psi}\,.
\end{equation}
As discussed in Sec.~\ref{suse:BsMix}, the CKM-induced couplings $\Gamma_{u,c}$ (see \eq{Gammau}) lead to additional constraints from $D_0-\bar{D}_0$ mixing. Since the impact of $\Gamma_b$ entering through $\Gamma_{u}$ and $\Gamma_c$ from \eq{Gammau} is suppressed by small CKM factors (${\cal O}(\lambda^{3})$ and ${\cal O}(\lambda^{2})$, respectively), the constraint from $D_0-\bar{D}_0$ mixing can be reduced in a scenario with $|\Gamma_b|>|\Gamma_s|$. The choice $|\Gamma_b|\sim 1$ and $|\Gamma_s|\sim 0.35$ saturates the bound from $B_s-\bar{B}_s$ mixing on the product $\Gamma_s^*\Gamma_b$, while it leads to a suppression by a factor $|V_{us}|^2|\Gamma_s|^2\sim 5\times 10^{-3}$ of $C_{D\bar D}$ with respect to $C_{B\bar B}$. The constraint on $C_{D\bar D}$ given in \eq{eq:boundCDD} is then automatically fulfilled once the constraint on $C_{B\bar B}$ from \eq{eq:boundCDD} is imposed.

In the case of the box contribution to $b\to s\mu^+\mu^-$, the coupling $\Gamma_\mu$ enters as an additional free
parameter, limited to values $\Gamma_\mu\lesssim\mathcal{O}(1)$ {in order} to ensure perturbativity. The 
$2\sigma$-bound from $B_s-\overline{B}_s$ mixing constrains $C_9^{\rm box}\;=\;-C_{10}^{\rm box}$  to
\begin{equation}
  |C_9^{\rm box}|\leq\;0.05\;\dfrac{\xi_9^{\rm box}}{\sqrt{\xi_{B\bar B}}}\; 
  |\Gamma_\mu|^2\;\dfrac{1\, {\rm TeV}}{m_\Psi}\,, 
\label{eq:c9boxbound}
\end{equation}
with the group factors $\xi_9^{\rm box}={{\chi}{\eta} - \chi^{\rm{M}}\eta^{\rm{M}}}$ given in Tab.~\ref{CBBC9factors}.
Considering the maximum value of the ratio  ${\xi_9^{\rm box}}/{\sqrt{\xi_{B\bar B}}}$, namely $4\sqrt{2/11}\simeq 1.7$ for
the representations C-I, C-II and D-I, D-II,  we find from \eq{eq:c9boxbound} that a solution of the $b\to s\mu^+\mu^-$ anomalies at the $2\sigma$-level requires a rather large coupling
\begin{equation}
\left| \Gamma_\mu\right|\geq 2.1\:\sqrt{\dfrac{m_\Psi}{1\, \rm TeV}}\,. 
\end{equation}

Let us now turn to the anomalous magnetic moment of the muon. In the limit of equal masses, the NP contribution is given by
\begin{equation}
  \Delta a_\mu=\;\pm(5.8\times 10^{-12})\;\xi_{a_\mu}\;|\Gamma_\mu|^2\;\left(\dfrac{1\, {\rm TeV}}{m_\Psi}\right)^2\,, 
\end{equation}
with $\xi_{a_\mu}=\chi_{a_\mu}(\eta_{a_\mu}-\widetilde\eta_{a_\mu})$ in Tab.~\ref{AMMfactors}. The plus applies to case a) while the minus applies to case b). In order to end up with a value for $a_\mu$ that falls within the experimental $2\,\sigma$ range, 
a positive NP contribution $\Delta a_\mu=6.2\times 10^{-10}$ is needed to have constructive interference with the SM.
This in turn implies the need for a positive (negative) group factor $\xi_{a_\mu}$ for case a) (b)), which can be accomplished for all representations by choosing an appropriate hypercharge $X\in[-1,+1]$. Selecting the representation C-II or C-V (C-I) and maximizing the effect in the anomalous magnetic moment by setting $X=1$ ($X=-1$), we find $\xi_{a_\mu}=16$ ($\xi_{a_\mu}=-24$) and that $a_\mu$ can be brought into agreement with the experimental measurement at the $2\sigma$-level for 
\begin{eqnarray}
\left| \Gamma_\mu\right|\geq 2.6 (2.1)\:\dfrac{m_\Psi}{\, \rm TeV}\,.
\end{eqnarray}

We see that both the tensions in $b\to s\mu^+\mu^-$ data and in the anomalous magnetic moment of the muon, $a_\mu$, 
can be reduced below the $2\,\sigma$ level for NP masses at the TeV scale and a coupling $|\Gamma_\mu|\ge 2.1$.
In light of this large value one might wonder, wether the LEP bounds on the $Z\mu^+\mu-$ coupling discussed in Sec.~\ref{sec:Zmumu} could become relevant. Evaluation of \eq{eq:Zmumu} gives
\begin{equation}
    \frac{\delta g_{L\,\mu}}{g_{L\,\mu}^{\textrm{SM}}}(m_Z^2)\;=\;-0.0006\%\;\xi_{Z}\;|\Gamma_\mu|^2\;\left(\dfrac{1\, {\rm TeV}}{m_\Psi}\right)^2\,, 
\end{equation}
with $\xi_{Z}=\chi_{Z}(\eta_{Z}/3+\widetilde\eta_{Z})$ in case a) and $\xi_{Z}=\chi_{Z}(\widetilde{\eta}_{Z}/3+\eta_{Z})$ in case b). For $|X|\leq1$, the group factor maximally reaches $\xi_Z\sim 10$
and the correction to the $Z\mu^+\mu^-$ vertex thus stays two orders of magnitude below the experimental sensitivity at LEP (see \eq{eq:Zmumu_LEP}) for masses of the new particle at the TeV scale.

In order to decide, whether a coupling $\Gamma_\mu$ of size $|\Gamma_\mu|\ge 2.1$ is still viable, it is further instructive to study the scale of the Landau pole of this coupling at the one-loop level.
This scale signals the break-down of the perturbative regime. Therefore, it provides an upper limit on the UV cut-off beyond which the theory needs to be complemented with new degrees of freedom if perturbativity 
shall be conserved. 
The Landau pole is obtained by evaluation of the renormalization-group equations (RGEs), which were determined at two loop for Yukawa couplings in a general quantum field theory e.g. in  Refs.~\cite{Machacek:1983fi,Luo:2002ti,Goertz:2015nkp}. 
For Yukawa-like couplings beyond the SM, the RGEs depend on the representations of the new particles under the SM gauge group. 
We studied the issue of the Landau pole for our models by implementing some of the possible scenarios in the public code {\sc Sarah}~\cite{Staub:2016dxq} and found that the running is dominated by $\mathcal{O}(\Gamma_\mu^2)$ corrections. For $|\Gamma_\mu|\le 2.4$, the respective terms in the RGE 
lead to a Landau pole at $\gtrsim10^3$ TeV.

In the case of $b\to s\mu^+\mu^-$, the requirement of a large coupling $|\Gamma_\mu|\ge 2.1$
is a consequence of the tight constraint from $B_s-\overline{B}_s$ mixing, and we will discuss in the following
the possibility to relax this constraint by considering non-degenerate masses for the new particles.

\begin{table}
\[
\begin{array}{*{20}{c}}
\xi_{a_\mu} &\vline& I&{II}&{III}&{IV}&V&{VI}\\
\hline
A&\vline&    2 \, X-1 & 2 \, X & \frac{3}{2}\, X-1 & \frac{1}{4} (6 \, X+1) &   2 \, X & \frac{3 }{2}\, X-1 \\
B&\vline&    6 \, X-3 & 6 \, X & \frac{9 }{2}\, X-3 & \frac{3}{4} (6 \, X+1) &   6 \, X & \frac{9}{2}\, X-3 \\
C&\vline&    16 \, X-8 & 16 \, X & 12 \, X-8 & 12 \, X+2 & 16 \, X & 12 \, X-8 \\
D&\vline&    6 \, X-3 & 6 \, X & \frac{9}{2}\, X-3 & \frac{3}{4} (6 \, X+1) &   6 \, X & \frac{9}{2}\, X-3 
\end{array}
\]
\caption{Group factors for the various representations entering the anomalous magnetic moment of the muon.}
\label{AMMfactors}
\end{table}
 
\subsection{Majorana case with non-degenerate masses}
\label{sec:uneqmass}

\begin{figure*}[t]
\begin{center}
\begin{tabular}{cp{7mm}c}
\includegraphics[width=0.6\textwidth]{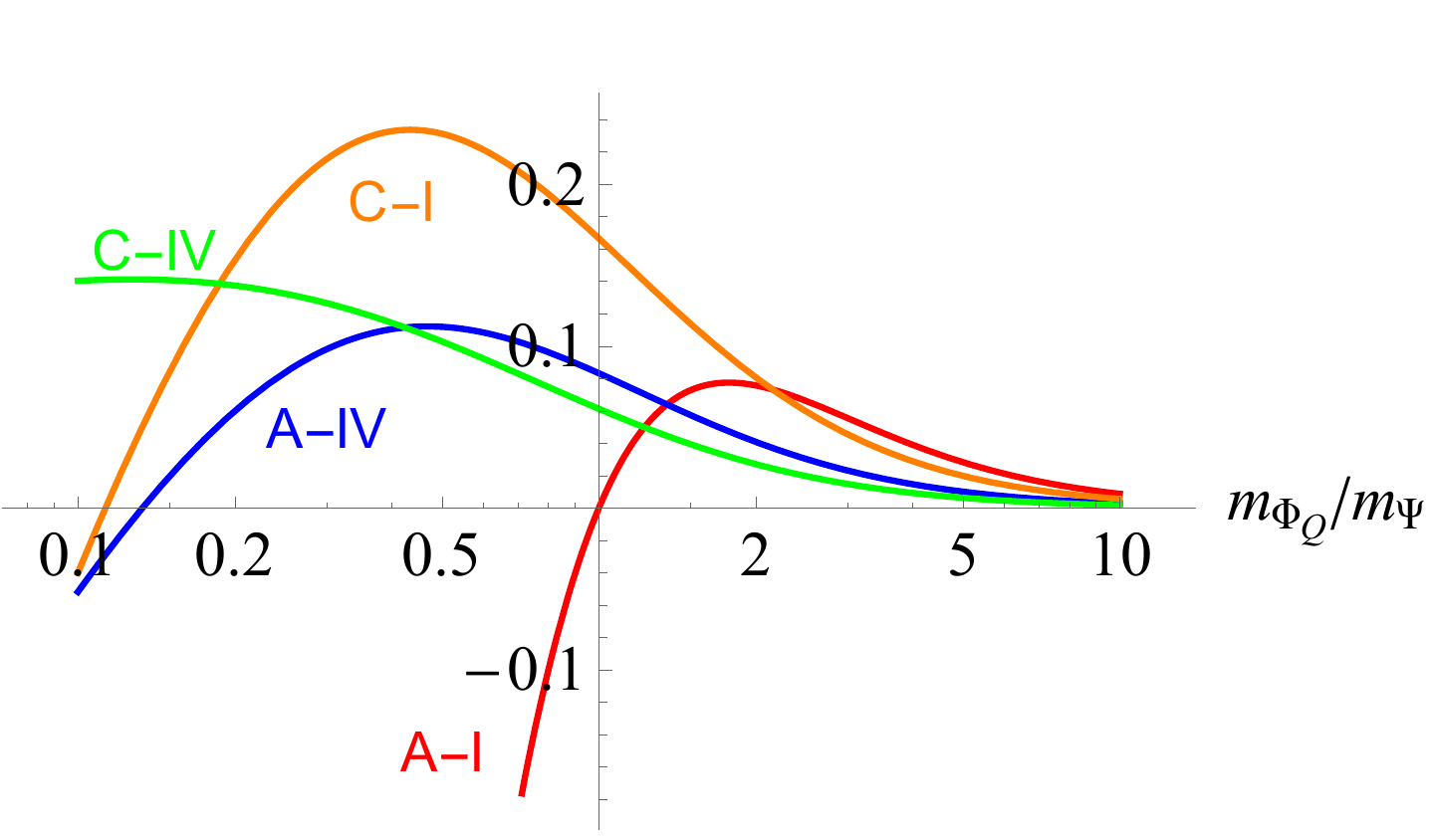}
\end{tabular}
\end{center}
\caption{The function $H^{a)}\left(m_{\Phi_Q}/m_\Psi  \right)$ entering $C_{B\bar B}$ for the four Majorana representations A-I, A-IV, C-I and C-IV. Note that representation C-IV only has a zero crossing for very small values of $m_{\Phi_Q}/m_\Psi$ outside the plot range.
 \label{CBBunequal} }
\end{figure*}

In this section, we address the question whether the impact of the constraint from 
$B_s-\overline{B}_s$ mixing can be reduced by considering a non-degenerate spectrum for the masses of the new particles. 
In the model classes a) and b), the Wilson coefficient $C_{B\bar B}$ for $B_s-\overline{B}_s$ mixing is proportional to the function
\begin{eqnarray}
   H^{a)}(m_{\Phi_Q}/m_\Psi)&=&{{\chi _{B\bar B}}{\eta _{B\bar B}}F\left( {x_Q,x_Q} \right) + 2\chi _{B\bar B}^{\rm{M}}\eta _{B\bar B}^{\rm{M}}G\left( x_Q,x_Q \right)},\nonumber\\
   H^{b)}(m_{\Psi_Q}/m_\Phi)&=&({{\chi _{B\bar B}}{\eta _{B\bar B}}-\chi _{B\bar B}^{\rm{M}}\eta _{B\bar B}^{\rm{M}})F\left( y_Q,y_Q \right)},
\end{eqnarray}
with $x_Q=m_{\Phi_Q}^2/m_\Psi^2$ and $y_Q=m_{\Psi_Q}^2/m_\Phi^2$. Note that both loop functions $F$ and $G$ have a smooth behavior with respect to their arguments and never switch sign. Therefore, 
a reduction of the effect in $B_s-\overline{B}_s$ mixing by varying the mass ratio
$m_{\Phi_Q}/m_\Psi$ or $m_{\Psi_Q}/m_\Phi$ is only possible through a (partial) cancellation between
the $F$- and $G$-term in the function $H$. Such a cancellation can only occur in the model class a) with the additional condition of $\Psi$ being a Majorana fermion because in all other cases only one loop-function $F$ is present.
Among the various representations, only four permit the Majorana option: A-I, A-IV, C-I and C-IV.
In Fig.~\ref{CBBunequal} we show the function $H^{a)}(m_{\Phi_Q}/m_\Psi)$ for these
four representations in the Majorana case. Each of the curves has a zero-crossing, given 
by $m_{\Phi_Q}/m_\Psi=1,\,0.11,\,0.13$ 
for A-I, A-IV and C-I, respectively, while it lies outside the plotted range for C-IV.

\begin{figure*}[t]
\begin{center}
\begin{tabular}{cp{7mm}c}
\includegraphics[width=0.6\textwidth]{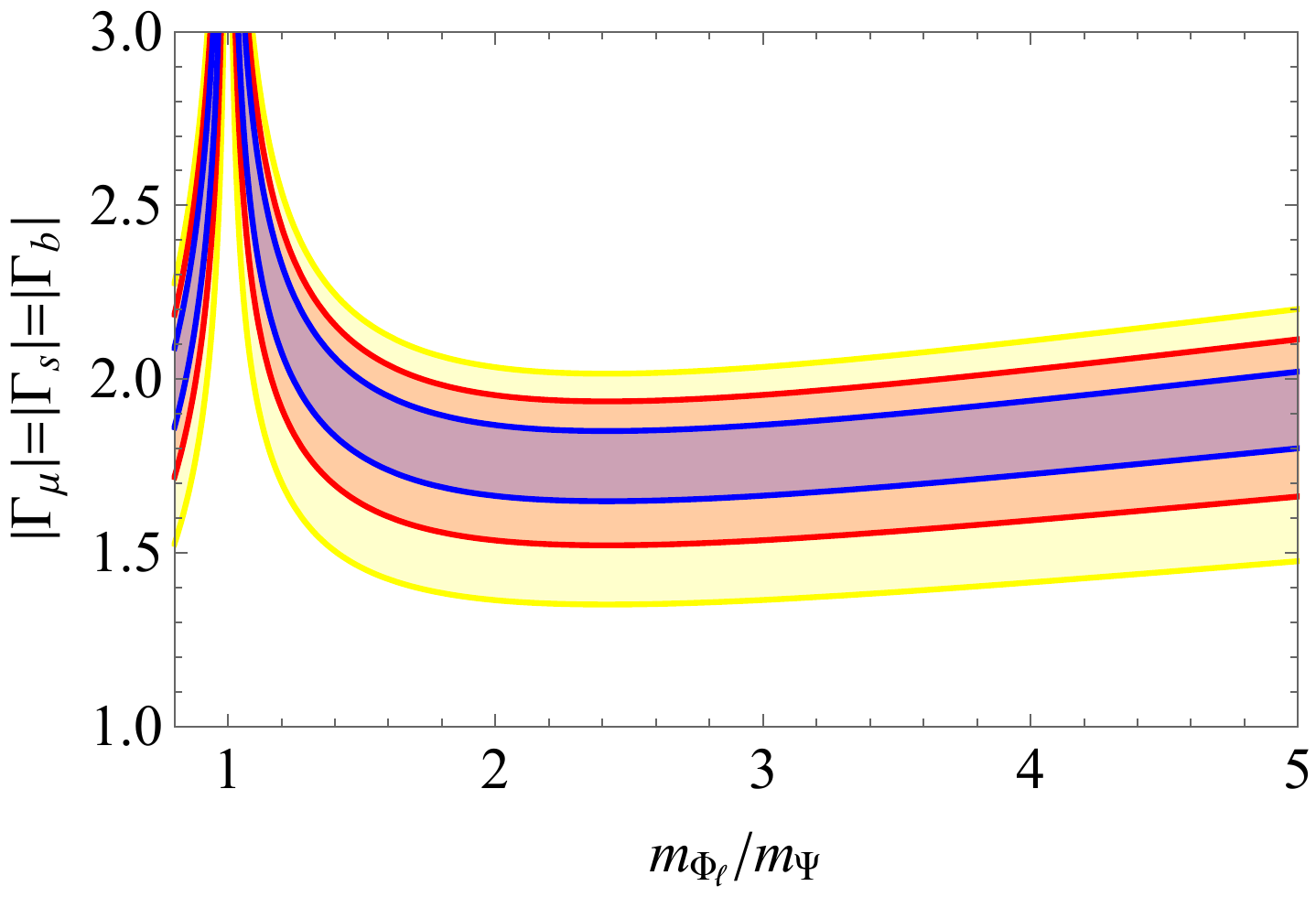}
\end{tabular}
\end{center}
\caption{Allowed regions for the coupling strength to muon, bottom and strange quarks from $b\to s\mu^+\mu^-$ data as a function of $m_{\Phi_\ell}/m_\Psi$ for case A-I in scenario a) with $m_{\Phi_Q}=m_\Psi=1\,$TeV. Blue, red and yellow correspond to $1\,\sigma$, $2\,\sigma$ and $3\,\sigma$, respectively.
 \label{GammaVSxL} }
\end{figure*}

Obviously, choosing a mass configuration that corresponds to the zero of of the function $H$ completely avoids 
any constraint from $\Delta F=2$ processes. Let us study the consequences for the representation A-I where
this situation occurs for $m_\Psi=m_{\Phi_Q}$. Note that the mass $m_{\Phi_\ell}$ of the scalar $\Phi_\ell$ has to be 
split from the one of the other two particles in order to get a non-vanishing contribution to $C_9^{\rm box}$. 
Under the simplifying assumption $|\Gamma_b|=|\Gamma_s|=|\Gamma_\mu|$, we show in Fig.~\ref{GammaVSxL} 
as a function of $m_{\Phi_\ell}/m_\Psi$ the generic coupling size needed to explain the $b\to s\mu^+\mu^-$ data.
We see that the larger space available in $\Gamma_s^*\Gamma_b$ in the absence of the bound from $B_s-\overline{B}_s$ mixing,
allows to obtain a solution at the $2\,\sigma$ level for a generic coupling size of $|\Gamma_b|=|\Gamma_s|=|\Gamma_\mu|\gtrsim 1.6$
for a mass splitting $m_{\Phi_\ell}/m_\Psi\gtrsim 2$ and $m_\Psi\sim 1\,$TeV. The Majorana property of
$\Psi$ constrains the photon penguin contribution because it fixes $q_{\Psi}=0$ and $q_{\Phi_Q}=-1/3$, leading to
\begin{equation}
   C_7\,=\,-C_9^\gamma\,=\,-0.005\,V_{ts}^*V_{tb}\,\left(\dfrac{1\, {\rm TeV}}{m_\Psi}\right)^2.
\end{equation}
For $|\Gamma_b|,|\Gamma_s|< 3$ and $m_\Psi= 1\,$TeV, we encounter values $|C_7|=|C_9^\gamma| < 0.044$, 
which are too small to have a relevant impact.

\begin{figure*}[t]
\begin{center}
\begin{tabular}{cp{7mm}c}
\includegraphics[width=0.6\textwidth]{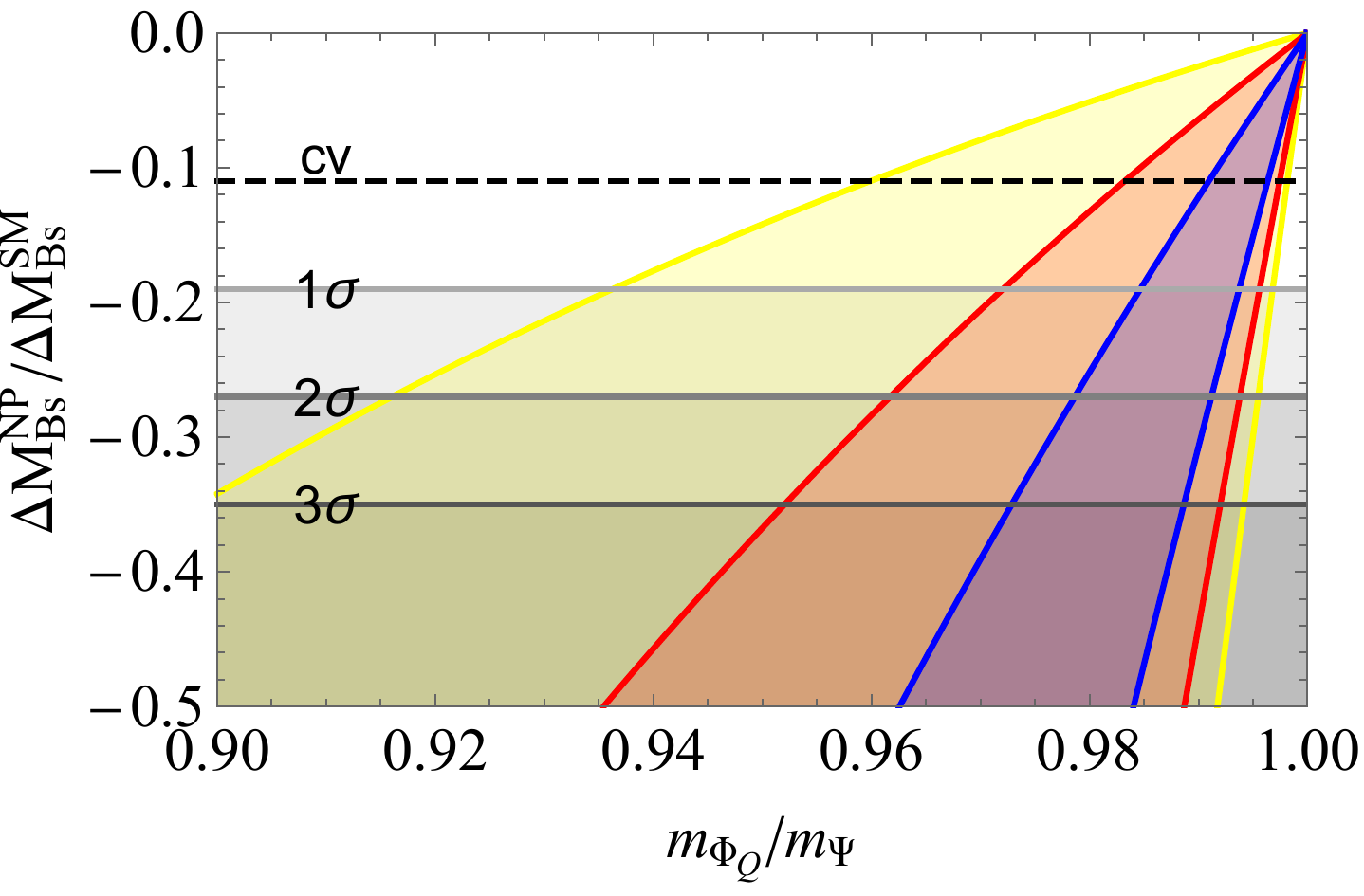}
\end{tabular}
\end{center}
\caption{Allowed regions in the $m_{\Phi_Q}/m_\Psi$--$\Delta M_{B_s}^{\rm
NP}/\Delta M_{B_s}^{\rm SM}$ plane from $b\to s\mu^+\mu^-$ data for $m_{\Phi_\ell}=2\,m_\Psi=2\,{\rm TeV}$ and $\Gamma_\mu=2$. Blue, red and yellow corresponds to $1\,\sigma$, $2\,\sigma$ and $3\,\sigma$, respectively. The regions below the gray lines are excluded by currant $B_s-\overline{B}_s$ mixing data at the denoted $\sigma$ level and the dotted line is the current central value.
 \label{xQvsDeltaMBs} }
\end{figure*}

In the case $m_{\Phi_Q}/m_\Psi<1$, a negative NP contribution to $\Delta M_{B_s}$ is generated, as preferred by current lattice data. This is illustrated in Fig.~\ref{xQvsDeltaMBs} where the effect on $\Delta M_{B_s}$ is shown as a function of $m_{\Phi_Q}/m_\Psi$, assuming that the $b\to s\mu^+\mu^-$ anomalies are accounted for by our model. An improvement in $B_s-\overline{B}_s$ mixing 
can be achieved simultaneously with a solution of the $b\to s\mu^+\mu^-$ anomalies if a small mass
splitting $0.98\lesssim m_{\Phi_Q}/m_\Psi\lesssim 1.0$ is introduced.


\section{Conclusions}
\label{sec:conclusions}

In this article we have studied the effects of new heavy scalars and fermions on $b\to s\mu^+\mu^-$ processes in a systematic way, aiming at an explanation of the observed deviations from the SM expectations. We investigated the two distinct cases of:
\begin{enumerate}[label=\alph*)]
\item one additional fermion $\Psi$ and two additional scalars $\Phi_Q$ and $\Phi_\ell$.
\item two additional fermions $\Psi_Q$ and $\Psi_\ell$ and one additional scalar $\Phi$.
\end{enumerate}
In both cases the additional particles interact with left-handed $b$-quaks, $s$-quarks and muons via Yukawa-like couplings $\Gamma_b$, $\Gamma_s$ and $\Gamma_\mu$,
respectively. Such a scenario is phenomenologically well motivated as it leads (to a good approximation) to 
the pattern $C_9=-C_{10}$ for the relevant Wilson coefficients, capable of improving the global agreement with 
$b\to s\mu^+\mu^-$ data by more than $4\,\sigma$. Considering representations up to the adjoint one under the SM gauge group, we classified all possible combinations of representations for the new particles that are allowed by charge conservation in the new Yukawa-type vertices  (24 for each case a) and b)). In this setup, we calculated the NP contributions to $b\to s\mu^+\mu^-$ processes, $B_s-\overline{B}_s$ mixing, $b\to s\gamma$, $b\to s\nu\bar\nu$ and the anomalous magnetic moment of the muon $a_\mu$, expressing the results in terms of loop functions times the group factors for the various representations (collected in tables).

In our numerical analysis we found that the constraints from $B_s-\overline{B}_s$ mixing are very stringent due to the new lattice data favoring destructive interference with the SM. In our models, the contributions to 
$B_s-\overline{B}_s$ mixing typically interferes constructively with the SM. A solution of the $b\to s\mu^+\mu^-$ anomalies at the $2\,\sigma$ level can only be obtained if a rather large muon coupling, $|\Gamma_\mu|\gtrsim 2.1$ for masses of the new particles
at the TeV scale, compensates the tight bounds on $\Gamma_s^*\Gamma_b$. The constraints from $B_s-\overline{B}_s$ mixing can be avoided in models of class a) under the additional condition of $\Psi$ being a Majorana fermion. Among the four representations that permit this situation, one features an exactly vanishing constribution to $B_s-\overline{B}_s$ mixing for degenerate masses $m_{\Phi_q}=m_\Psi$. For this representation, the $b\to s\mu^+\mu^-$ data can be accounted for at the $2\,\sigma$ level with a coupling size $|\Gamma_b|=|\Gamma_s|=|\Gamma_\mu|\gtrsim 1.6$ if $m_{\Phi_q}\gtrsim m_\Psi$. The contribution that our models generate to the anomalous magnetic moment of the muon $a_\mu$ only depend on the muon coupling $\Gamma_\mu$. An explanation of the long-standing anomaly in $a_\mu$ at the $2\,\sigma$ level again requires rather large values $|\Gamma_\mu|\gtrsim 2.1$ for this coupling, requiring the presence of additional new particles at a scale $\lesssim 10^3$ TeV or below in order to guarantee perturbativity of the theory.

As our model with the minimal number of new particles (three) gives rise to a $C_9=-C_{10}$ solution for 
$b\to s\mu^+\mu^-$ data, $B_s\to\mu^+\mu^-$ is predicted to be below SM expectations. Therefore, a SM-like branching ratio for $B_s\to\mu^+\mu^-$ would lead to the requirement of more then three new particles in 
order to explain the $b\to s\mu^+\mu^-$ anomalies via a loop effect involving heavy fermions and scalars.

 \acknowledgments{
 We are grateful to Bernat Capdevila for providing us the updated $2\sigma$ range of the the $C_9=-C_{10}$ fit from Ref.~\cite{Descotes-Genon:2015uva}. The work of A.~C. has been supported by a Marie Curie Intra-European Fellowship of the European Community's 7th Framework Programme under contract number PIEF-GA-2012-326948 and by an Ambizione Fellowship of the Swiss National Science Foundation. P.~A., L.~H. and F.~M. acknoledge the financial support from FPA2013-46570, 2014-SGR-104, and 
project MDM-2014-0369 of ICCUB (Unidad de Excelencia `Maria de Maeztu'). L.~H. and F.~M. have further been supported 
by project FPA2014-61478-EXP. A.C. thanks the Aspen Center for Physics for hospitality during completion of this work.}

\bibliography{BIB}

\end{document}